\documentclass[aps,pre,superscriptaddress,twocolumn,floatfix,showkeys,showpacs]{revtex4}

\usepackage{amsmath,bm,graphicx}
\def \ed {\end{document}}
\def\Fbox#1{\vskip1ex\hbox to 8.5cm{\hfil\fboxsep0.3cm\fbox{%
  \parbox{8.0cm}{#1}}\hfil}\vskip1ex\noindent}  %%  {TEXT} in BOX

\begin{document}

\title{Reconnection Dynamics for Quantized Vortices}

\thanks{The authors of this manuscript would not be studying this subject were it not for the influence and collaboration of K.R. Sreenivasan, for which we are profoundly grateful.}

\author{M. S. Paoletti}
\affiliation{Department of Physics, and Institute for Research in Electronics and Applied Physics, University of Maryland, College Park, MD 20742}
\author{Michael E. Fisher}
\affiliation{Institute for Physical Science and Technology, University of Maryland, College Park, MD 20742}
%\thanks{}
\author{D. P. Lathrop}
\email[email address: ]{lathrop@umd.edu}
\affiliation{Department of Physics, and Institute for Research in Electronics and Applied Physics, University of Maryland, College Park, MD 20742}
\affiliation{Institute for Physical Science and Technology, University of Maryland, College Park, MD 20742}
\affiliation{Department of Geology, University of Maryland, College Park, MD 20742}

\begin{abstract}
By analyzing trajectories of solid hydrogen tracers in superfluid $^4$He, we identify tens of thousands of individual reconnection events between quantized vortices.  We characterize the dynamics by the minimum separation distance $\delta(t)$ between the two reconnecting vortices both before and after the events.  Applying dimensional arguments, this separation has been predicted to behave asymptotically as $\delta(t) \approx A \left ( \kappa \vert t-t_0 \vert \right ) ^{1/2}$, where $\kappa=h/m$ is the quantum of circulation.  The major finding of the experiments and their analysis is strong support for this asymptotic form with $\kappa$ as the dominant controlling feature, although there are significant event to event fluctuations.  At the three-parameter level the dynamics may be about equally well-fit by two modified expressions: (a) an arbitrary power-law expression of the form $\delta(t)=B \vert t-t_0 \vert ^{\alpha}$ and (b) a correction-factor expression $\delta(t)=A\left (\kappa \vert t-t_0 \vert \right ) ^{1/2}(1+c \vert t-t_0 \vert )$.  The measured frequency distribution of $\alpha$ is peaked at the predicted value $\alpha=0.5$, although the half-height values are $\alpha=0.35$ and 0.80 and there is marked variation in all fitted quantities.  Accepting (b) the amplitude $A$ has mean values of $1.24 \pm 0.01$ and half height values of 0.8 and 1.6 while the $c$ distribution is peaked close to $c=0$ with a half-height range of $-0.9$~s$^{-1}$ to 1.5~s$^{-1}$.  In light of possible physical interpretations we regard the correction-factor expression (b), which attributes the observed deviations from the predicted asymptotic form to fluctuations in the local environment and in boundary conditions, as best describing our experimental data.  The observed dynamics appear statistically time-reversible, which suggests that an effective equilibrium has been established in quantum turbulence on the time scales ($\leq 0.25$~s) investigated.  We discuss the impact of reconnection on velocity statistics in quantum turbulence and, as regards classical turbulence, we argue that forms analogous to (b) could well provide an alternative interpretation of the observed deviations from Kolmogorov scaling exponents of the longitudinal structure functions.
\end{abstract}

\pacs{insert PACS here}

\maketitle

% main text
\section{Introduction}
\label{introduction}

Relaxation toward equilibrium requires dissipative processes but can be inhibited by topological defects that cannot diffuse.  Linear topological defects occur in a variety of systems such as superfluids \cite{donnelly91}, liquid crystals \cite{chuang91} and superconductors \cite{blatter94}. Dissipation normally accompanies the reconnection of two defect lines that cross, change topology by exchanging ends and separate (as illustrated in Fig.\ 1). Prime examples \cite{priest00} of dissipation by reconnection occur in astrophysical plasmas (such as solar flares \cite{lin71, lin03} and magnetic substorms \cite{terasawa76, baker76}) and sawtooth crashes in fusion devices \cite{savrukhin01}, where magnetic energy is dissipated by the acceleration of nearby particles \cite{oieroset02, dmitruk03, holman03,drake05,drake06}. Reconnection has also been studied in liquid crystals \cite{chuang91}, superconductors \cite{brandt91,blatter94,bou-diab01}, cosmic strings \cite{Hindmarsh95}, viscous \cite{fohl75,ashurst87, kerr89} and Euler \cite{siggia85} vortices, Bose-Einstein condensates \cite{caradoc00} and superfluids \cite{feynman55, schwarz85, schwarz88, koplik93, deWaele94, gabbay98, lipniacki00, kivotides01, leadbeater01, vinen01, ogawa02, nazarenko03, vinen05, kuzmin06, bewley08, paoletti08a}.

\begin{figure}[t]
\begin{center}
\includegraphics[width=.48\textwidth]{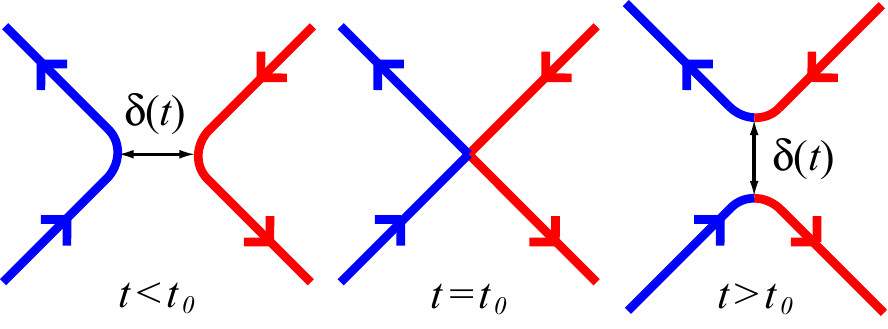}
\caption[Reconnection schematic and example]{Schematic of the evolution of reconnecting antiparallel directed linear topological defects; in the case of vortices the arrows indicate the sense of the vorticity.  The minimum separation between the defects is $\delta(t)$ with $\delta(t_0)=0$.  Although not evident here, reconnection need not be planar and, indeed, has been predicted to be intrinsically three-dimensional \cite{schwarz85, deWaele94}.}
\label{reconschem}
\end{center}
\end{figure}

Superfluid $^4$He behaves as a mixture of two interpenetrating fluids, a viscous normal fluid and an inviscid superfluid \cite{donnelly91}.
Vorticity in the superfluid component is confined to linear topological defects, called quantized vortices, which possess circulation values that are integral multiples of $\kappa = h/m =9.97\times10^{-4}$ cm$^2/$s, where $h$ is Planck's constant and $m$ is the mass of a helium atom.  Viscosity does not lead to diffuse vorticity in a quantum fluid; this is clearly different than in classical fluids.  Superfluid $^4$He driven far from equilibrium consists of a complex tangle of quantized vortices, known as quantum turbulence \cite{feynman55}. Quantized vortex reconnection reduces the defect line length and transfers the energy to the normal fluid via mutual friction \cite{vinen57a, vinen57b, vinen57c} or into acoustic and Kelvin wave emission \cite{leadbeater01, vinen01, ogawa02, vinen05, kuzmin06}. This dissipative process is particularly significant in the limit of absolute zero temperature owing to the vanishing viscosity of the normal fluid component \cite{walmsley07}.

In this article, we characterize experimental data for nearly 20,000 quantized vortex reconnection events in superfluid $^4$He visualized using micron-sized solid hydrogen tracer particles.  Previous studies have shown that hydrogen tracers can be trapped by quantized vortices \cite{poole05, bewley06} in superfluid $^4$He and thereby used to directly visualize quantized vortex reconnection \cite{bewley08, paoletti08a}.  We characterize the dynamics of reconnection by measuring the minimum separation distance $\delta(t)$ between the vortices both before and after the reconnection event (see Fig.\ \ref{reconschem}).  Each measured separation sequence has been fit to two expressions each described by three parameters, namely, an arbitrary power-law of the form
\begin{equation}
\delta(t)=B \vert t-t_0 \vert ^{\alpha},
\label{alphafit}
\end{equation}
and a correction-factor expression
\begin{equation}
\delta(t)=A\left (\kappa \vert t-t_0 \vert \right ) ^{1/2} \left (1+c \vert t-t_0 \vert \right ).
\label{fisherfit}
\end{equation}
The two expressions suggest distinct physical interpretations and implications that will be a focus of this manuscript.  We provide a discussion of relevant previous work in Section~\ref{Previous}.  The details of the experiments and our techniques for identifying reconnection events are discussed in Section~\ref{Experiments}.  The arbitrary power-law expression (\ref{alphafit}) and the correction-factor expression in (\ref{fisherfit}) are discussed and compared in Section~\ref{Expressions}.  We present arguments for statistical time-reversibility and anisotropy in Section~\ref{Time-Reversibility}.  The effects of reconnection on quantum turbulence are discussed in Section~\ref{QT}.  An overview of potential implications for understanding classical turbulence is presented in Section~\ref{CT} while Section~\ref{Conclusions} summarizes our conclusions and suggests future investigations.

\section{Previous Studies}
\label{Previous}

Quantized vortex reconnection has been previously studied numerically and analytically by employing vortex-line methods \cite{schwarz78, schwarz85, schwarz88, deWaele94, lipniacki00, kivotides01} and by integrating the Gross-Pitaevskii equation \cite{koplik93, nazarenko03}.  To characterize the evolution of reconnecting vortices, some of these theoretical studies examined the minimum separation distance $\delta(t)$ between the vortices as shown in Fig.\ 1.  Assuming that the only relevant parameter in reconnection dynamics is the quantum of circulation $\kappa$, dimensional analysis yields the relation
\begin{equation}
\delta(t)=A\left (\kappa\vert t-t_0 \vert \right ) ^{1/2},
\label{delta}
\end{equation}
where $t_0$ is the reconnection moment for the vortices and $A$ is a dimensionless factor of the order unity. One may optimistically expect this scaling to be valid for length scales between the vortex core size ($\sim 1$~\AA) and the typical intervortex spacing ($\sim 0.1-1$~mm for our work here).  Basically, however, (\ref{delta}) should represent an asymptotic expression, subject at least to corrections for longer times.  Indeed, slight deviations were observed in prior simulations implementing line-vortex models \cite{schwarz85, deWaele94}.  The Gross-Pitaevskii equation is Hamiltonian and time-reversal invariant, but particular solutions may well break this symmetry.

Recent experimental studies have demonstrated that hydrogen tracer particles may be used to directly visualize quantized vortex reconnection \cite{bewley08, paoletti08a}.  Such tracer particles can be trapped on quantized vortices or may move with the normal fluid under the influence of Stokes drag \cite{poole05, bewley06, paoletti08b}.  In \cite{bewley08, paoletti08a} it was found that the form (\ref{delta}) must be modified to adequately represent the set of experimental data.  Specifically, in \cite{bewley08} the separation $\delta(t)$ for $t>t_0$ for 52 distinct events was fit to the form (\ref{alphafit}) while in \cite{paoletti08a} the alternative form (\ref{fisherfit}) was used both before and after approximately 20,000 reconnection events.  Here we undertake a comparison of these two different expressions.

\section{Experiments and Techniques}
\label{Experiments}

\subsection{Pulsed Counterflow Experiments}
\label{PulsedCF}

Our experiments are conducted in a cylindrical cryostat with a 4.5 cm diameter containing liquid $^4$He. The long axis of the channel is vertical with four 1.5 cm windows separated by 90$^{\circ}$.  A room-temperature mixture of 2\% H$_2$ and 98\% $^4$He by partial-pressure is injected into liquid helium \cite{bewley06} above the superfluid transition temperature ($T_{\lambda}=2.17$~K).  The injected hydrogen freezes producing a polydisperse distribution of tracer particles with diameters near 1~$\mu$m and an initial volume fraction of approximately $10^{-6}-10^{-5}$.  We subsequently cool the fluid evaporatively to the desired temperature in the range 1.70~K $< T <$ 2.05~K.  A portion of the hydrogen leaves the observation volume resulting in a volume fraction $\sim 10^{-7}$.  The solid hydrogen particles are illuminated by an argon ion laser sheet that is 8~mm high and 100~$\mu$m thick and has an optical laser power between 2 and 6~W.  A camera gathers 90$^{\circ}$ scattered light with a resolution of 16~$\mu$m per pixel at either 60, 80, or 100 frames per second.

A quantum vortex tangle and its accompanying cascade of reconnection events is induced by reproducibly driving the system away from equilibrium by a thermal counterflow \cite{donnelly91, paoletti08a, paoletti08b}, which increases the total vortex line length present in the system \cite{vinen57a, vinen57b, vinen57c, schwarz88}.  Upon cessation of the counterflow, we acquire data while the system relaxes toward equilibrium \cite{skrbek03}.  A spiral nichrome wire heater located at the bottom of the channel 7.5~cm below the observation volume initiates the counterflow. A fixed heat flux in the range 0.064~W/cm$^2<q<0.17$~W/cm$^2$ drives the system for approximately 5~s, after which it is allowed to relax for approximately 10~s before repeating the process. We employ a two-dimensional particle-tracking algorithm \cite{weeks} with sub-pixel precision to obtain single particle trajectories.  We estimate the uncertainty in the particle positions to be less than 4~$\mu$m.  While superficially similar, this technique is distinct from particle image velocimetry (PIV) \cite{adrian84}.  PIV obtains velocities by computing cross correlations of groups of particles, thereby requiring a smoothly-varying velocity field.  Given the two-fluid nature of superfluid helium this technique would be unsuitable for the present studies.

\begin{figure}[t]
\begin{center}
\includegraphics[width=.48\textwidth]{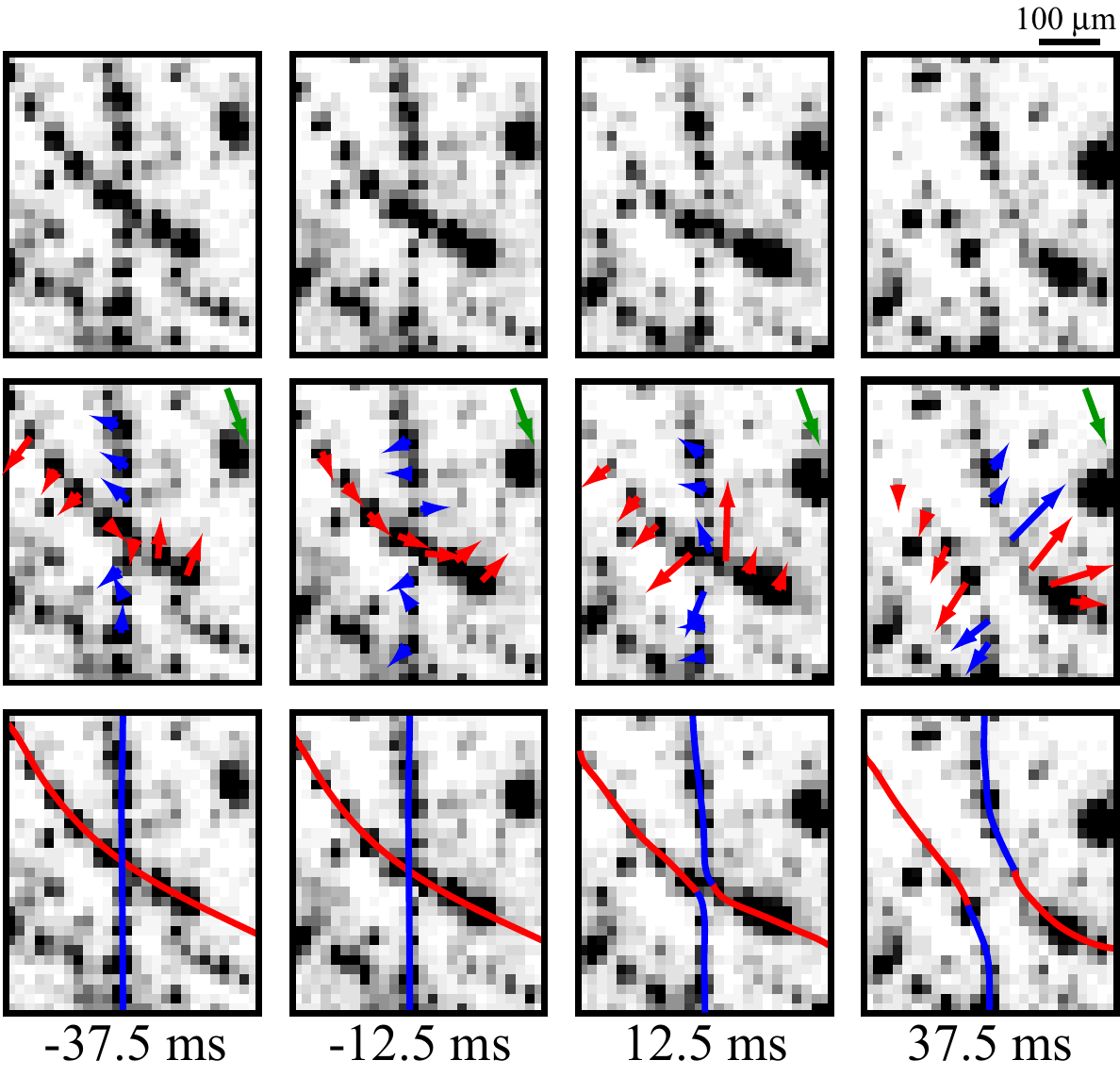}
\caption{Contrast-enhanced negative images of particles trapped on reconnecting vortices (top) along with velocity vectors (middle) and our interpretation of the pre- and post-reconnection configurations of the vortices denoted by the red and blue lines
(bottom) with time measured from $t_0$. The green vectors show the background drift that is subtracted from all velocity vectors. The
red and blue velocity vectors (middle) correspond to the red and blue marked vortices in the bottom images. The volume fraction of
hydrogen in these images ($10^{-5}$) is higher than for all the pulsed counterflow experiments discussed below. Reconnection is
particularly unambiguous in the online movies in \cite{paoletti08a}.}
\label{reconimages}
\end{center}
\end{figure}

\subsection{Identifying Reconnection Events}
\label{Identifying}

Near the reconnection moment $t_0$, reconnecting vortices move with high, atypical velocities and accelerations.  An example of a reconnection event is shown in Fig.\ \ref{reconimages} (also see Fig.\ 2 in \cite{bewley08} and note that reconnection is clearly evidenced in the online movies in \cite{paoletti08a}).  In this example, the particle density is high so that both vortices are marked by multiple trapped hydrogen particles.  The two vortices merge, exchange tails, then separate as indicated by the velocity vectors in the middle row of images in Fig.\ \ref{reconimages}.

Since the hydrogen particles are not completely passive, the hydrogen volume fraction in the pulsed counterflow experiments presented here has typically been kept one to two orders of magnitude lower than that shown in Fig.\ \ref{reconimages}.  For such low volume fractions, each identified vortex has only one to a few hydrogen particles trapped, thereby minimizing the effects of the hydrogen on the reconnection dynamics \cite{paoletti08b}.  A reconnection event is characterized, then, by a pair of particles rapidly approaching or separating.  The number of possible particle pairs analyzed is $\sim 10^{10}$, which requires an \textit{ad hoc} criterion to determine likely reconnection events.  We define particles $i$ and $j$ as marking a reconnection event at time $t$ if the pairwise separation $\delta_{ij}(t) = \vert \mathbf{r}_i(t)-\mathbf{r}_j(t) \vert$ satisfies
\begin{equation}
\xi_{ij} \equiv \delta_{ij}(t \pm 0.25 \mathrm{\;s})/\delta_{ij}(t)>\mathrm{4,}
\label{deltacrit}
\end{equation}
where $\mathbf{r}_i(t)$ is the two-dimensional projection of the position of particle $i$ at time $t$ and the plus (minus) sign indicates particles that separated after (approached before) an event, which we label as forward (reverse) events.  We choose the temporal duration of 0.25~s to allow a sufficient range to perform the power-law fits to the data while curtailing greater times, which are dominated by boundary effects and the presence of neighboring vortices.  The criterion (\ref{deltacrit}) excludes all but a fraction of possible pairs, namely $\sim 5\times 10^4$ forward and a similar number of reverse events.

\begin{figure}[t]
\begin{center}
\includegraphics[width=.48\textwidth]{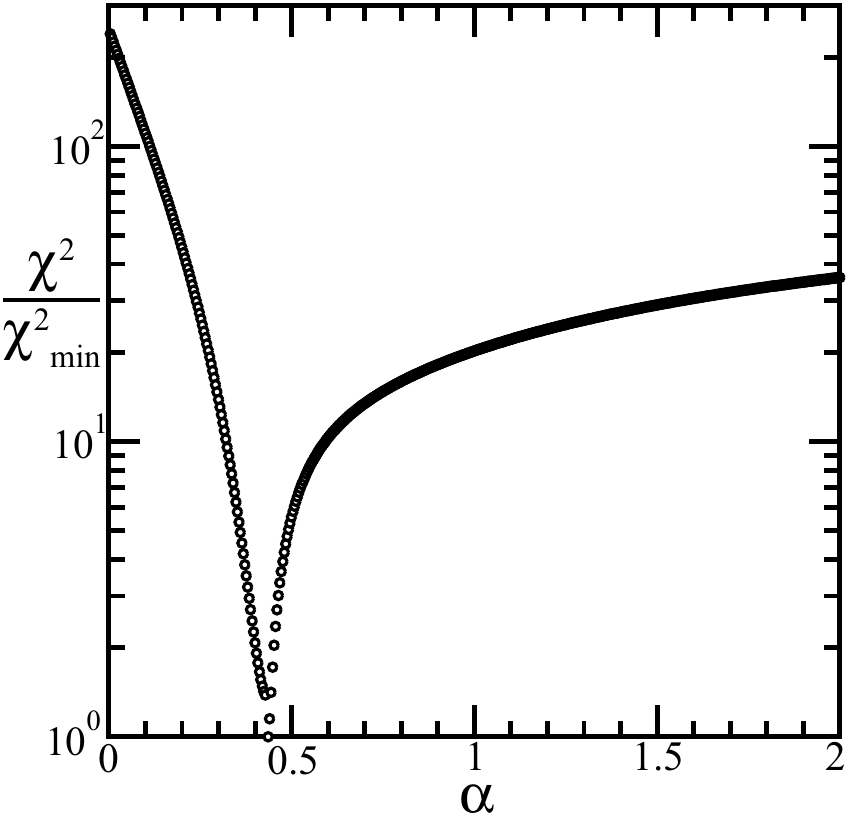}
\caption{Variation of $\chi^2(\alpha)$ normalized by its minimum value $\chi^2_{\mathrm{min}}=0.73$ as a function of the scaling exponent $\alpha$ for the event shown by the red squares in Fig.\ \ref{correction_events} below.  We choose the parameters of the arbitrary power-law fit \{$\alpha$, $B$, $t_0$\} that minimize $\chi^2$  as a function of $\alpha$ defined by (\ref{chi2eqn}).}
\label{chi2vsalpha}
\end{center}
\end{figure}

It is important to note that we are assuming $\delta(t) \simeq \delta_{ij}(t)$; however, the particles ($i$, $j$) may not be located as close as desirable to the point of reconnection.  We do not observe any correlations between the measured quantities discussed below and the initial particle separations or the values of $\xi_{ij}$ as defined in (\ref{deltacrit}); nevertheless, more detailed theoretical analyses of vortex reconnection are needed to reveal and quantify systematic effects that may be caused by interpreting our measurements of $\delta_{ij}(t)$ as good approximations to $\delta(t)$.

\section{Reconnection Dynamics}
\label{Expressions}

\subsection{Arbitrary Power-Law}
\label{Power-Law}

\begin{figure}[b]
\begin{center}
\includegraphics[width=.48\textwidth]{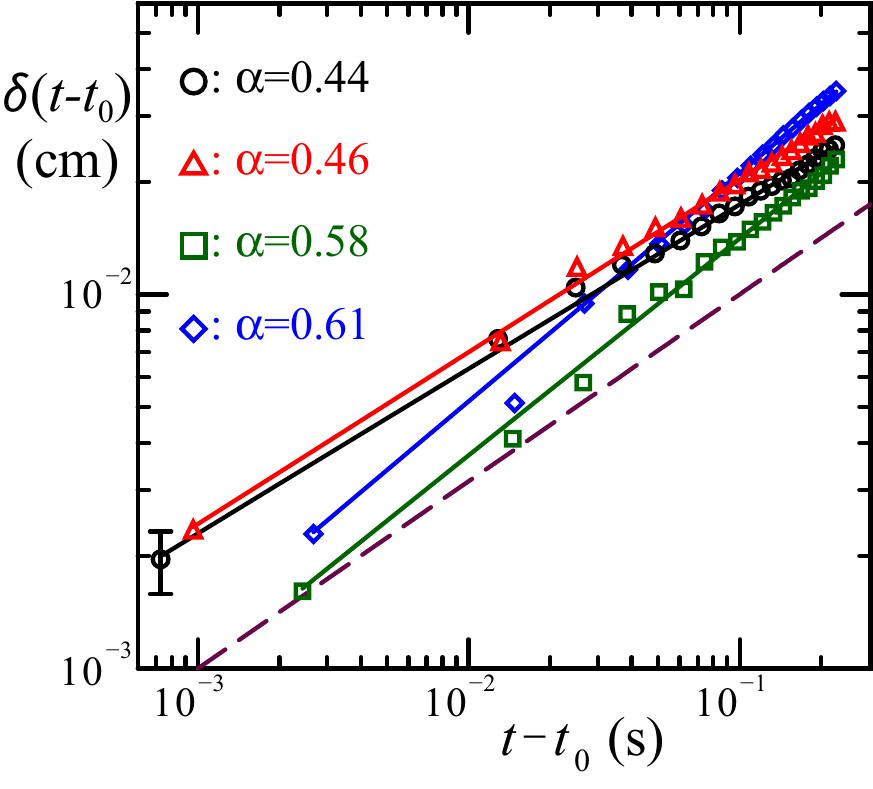}
\caption{Four forward events well fit by the arbitrary power-law expression (\ref{alphafit2}).  Symbols denote the measured separation $\delta(t)$ of pairs of particles on reconnecting vortices with an example error bar $\sigma=4$ $\mu$m while solid lines show fits to $\delta(t)=B\vert t-t_0\vert ^{\alpha}$ with $\alpha$ given in the legend.  The predicted asymptotic scaling $\delta(t) =(\kappa\vert t-t_0\vert)^{1/2}$ is shown by the purple dashed line. }
\label{alpha_events}
\end{center}
\end{figure}

\begin{figure}[t]
\begin{center}
\includegraphics[width=.48\textwidth]{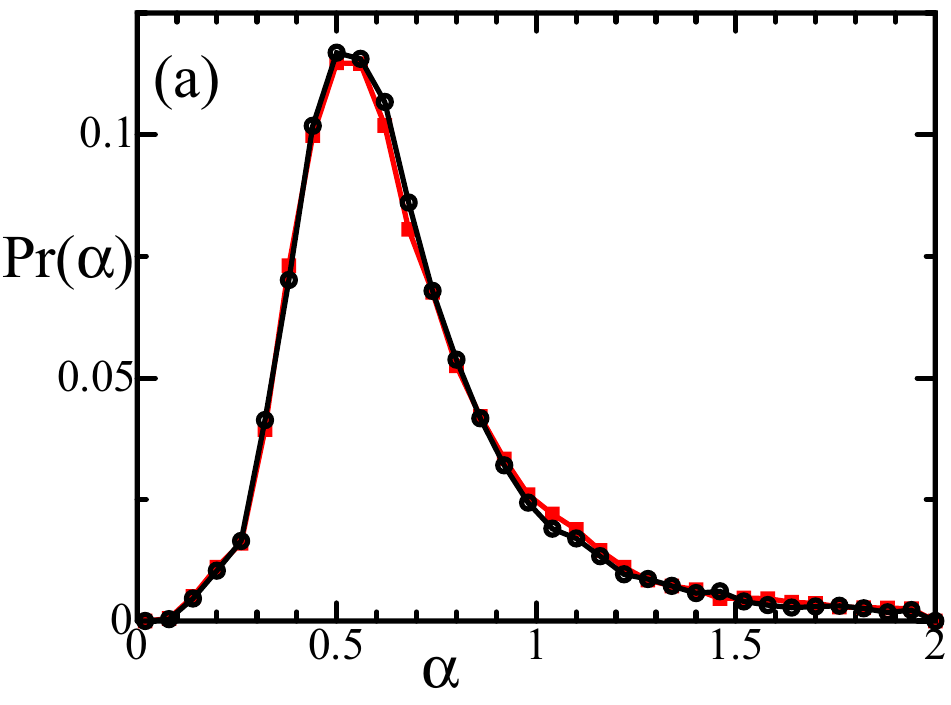}
\includegraphics[width=.48\textwidth]{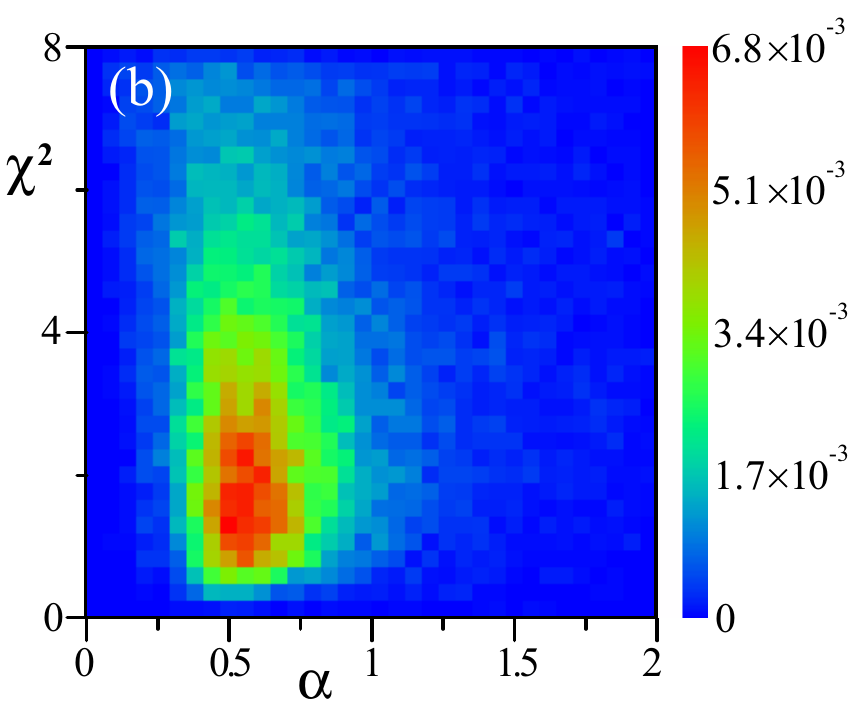}
\caption{(a) Normalized frequency distributions of $\alpha$ computed for 19,150 forward events (black circles) and 18,900 reverse events (red
squares). The mean values of $\alpha$ for forward and reverse events are 0.68 and 0.69, respectively.  (b) Two-dimensional contour diagram of $\chi^2$ versus $\alpha$ for forward events.  The peak near $\alpha=0.5$ with low values of $\chi^2$ indicates that (\ref{alphafit2}) best describes events with dynamics near those predicted in (\ref{delta}).}
\label{alpha_dists}
\end{center}
\end{figure}

We characterize the dynamics of reconnection by measuring the separation $\delta(t) \simeq \delta_{ij}(t)$ of pairs of particles ($i$, $j$) that meet the criterion (\ref{deltacrit}).  As mentioned above, previous dimensional and theoretical arguments predict that $\delta(t)$ behaves asymptotically as a power-law with a scaling exponent $\alpha=0.5$.  To test this hypothesis we fit our data to an arbitrary power-law of the form
\begin{equation}
\delta(t)=B \vert t-t_0 \vert ^{\alpha}.
\label{alphafit2}
\end{equation}
The values of $B$ and $t_0$ are determined by a linear least-squares fit of $[\delta(t)]^{1/\alpha}$ for 500 values of $\alpha$ evenly-spaced in the interval $0<\alpha<2$.  For each set of \{$\alpha$, $B$, $t_0$\} we compute the error in the fit
\begin{equation}
\chi^2(\alpha) \equiv \frac{1}{n} \sum_{m=1}^n \left [\frac{\delta^{\rm{fit}}_m-\delta_m}{\sigma} \right ]^2,
\label{chi2eqn}
\end{equation}
where $m$ denotes the movie frame, $\sigma =4$ $\mu$m (0.25 pixels) is an estimate of the uncertainty of the particle positions, and $n=$ 15, 20, 25 for data collected at 60, 80, or 100 frames per second, respectively.  We then choose the set of \{$\alpha$, $B$, $t_0$\} that minimizes $\chi^2$ (see Fig.\ \ref{chi2vsalpha}).

\begin{figure}[t]
\begin{center}
\includegraphics[width=.48\textwidth]{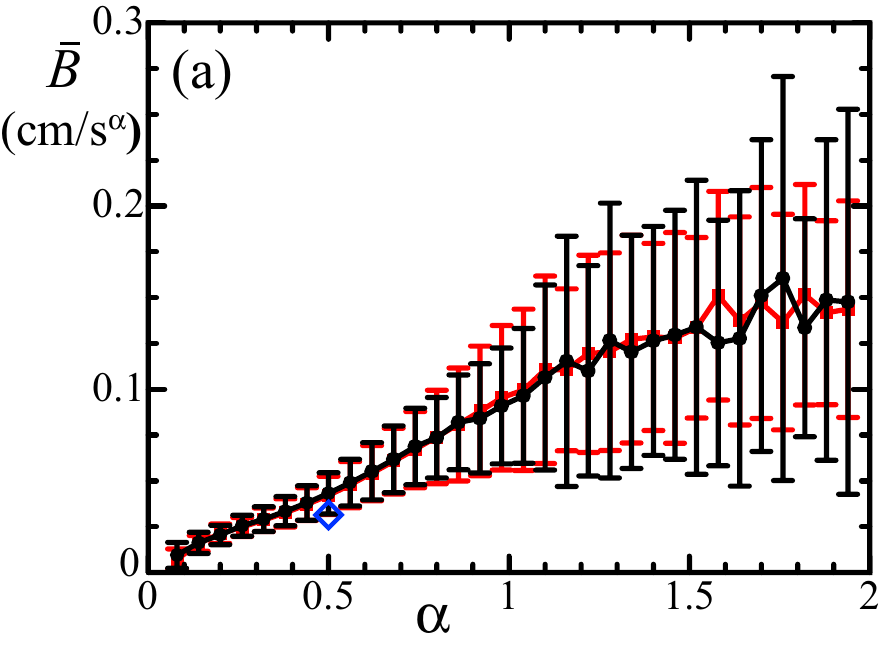}
\includegraphics[width=.48\textwidth]{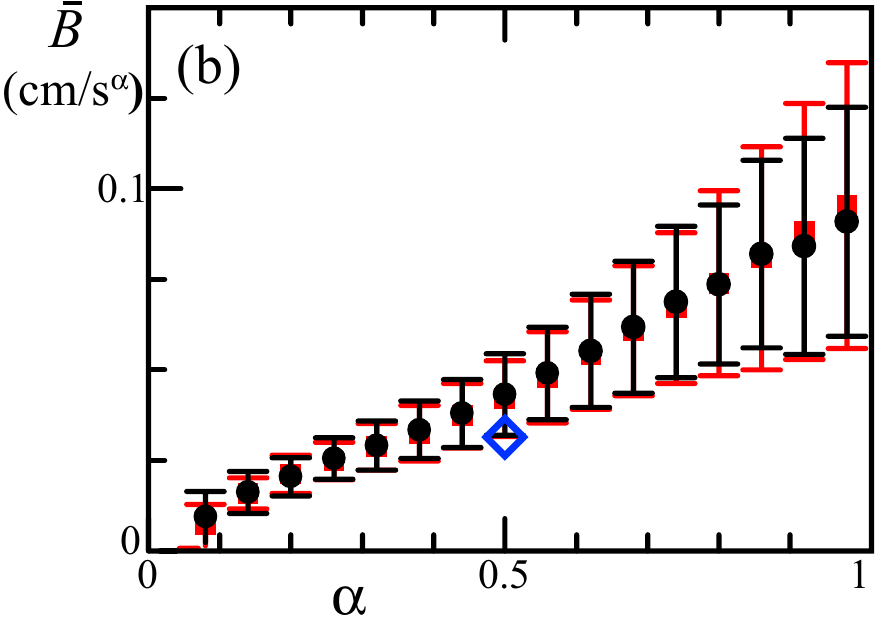}
\caption{Plots of the average fitted amplitude $\bar{B}$ as correlated with the scaling exponent $\alpha$ for 19,150 forward events (black circles) and 18,900 reverse events (red squares).  The error bars indicate the standard deviation of the data within the given range of $\alpha$.  The blue diamonds at $\alpha=0.5$ show $\bar{B}=\kappa^{1/2}$ to compare to the predicted scaling.  (a) The entire range of $\alpha$; (b) range restricted to $0<\alpha<1$.}
\label{B_dists}
\end{center}
\end{figure}

The measured separations $\delta(t)$ for four forward events are shown in Fig.\ \ref{alpha_events}, along with the predicted asymptotic form $\delta(t)=(\kappa \vert t-t_0 \vert )^{1/2}$ for comparison.  Fits to (\ref{alphafit2}) are shown as solid lines with the scaling exponent $\alpha$ given in the legend.  The most frequent fitted exponents cluster around the predicted value of $\alpha=0.5$ and their corresponding amplitudes $B$ are of order $\kappa^{1/2}$; however, there is a broad spread in both quantities.

Distributions of $\alpha$ for both the forward and reverse events, determined from fifty distinct experimental heat pulses, are shown in Fig.\ \ref{alpha_dists}(a).  The distributions are formed from events with $\chi ^2 < 4$. Approximately 40\% of the 50,000 pairs that meet the criterion in (\ref{deltacrit}) meet this $\chi ^2$ criterion.  Both distributions are asymmetric but peaked within 10\% of the predicted value $\alpha=0.5$.  Furthermore, as shown in Fig.\ \ref{alpha_dists}(b), events with fitted values near 0.5 typically have lower values of $\chi^2$.

The amplitudes $B$ for the same events are strongly correlated with the scaling exponent $\alpha$ as shown in Fig.\ \ref{B_dists}.  We find that events with $\alpha \simeq 0.5$ have amplitudes $B \simeq \sqrt{\kappa}$, as expected from dimensional analysis.  However, de Waele and Aarts \cite{deWaele94} measured $B \simeq \sqrt{\kappa/2\pi}$ in numerical simulations of quantized vortex reconnection in superfluid $^4$He using line-vortex methods; this is approximately 30\% of our experimentally determined value.  The time-scales in our experiments differ greatly from these numerical simulations; de Waele and Aarts determined their value of $B$ for $0<t_0-t<3$~$\mu$s, whereas our time-scales span 1~ms~$<\vert t-t_0 \vert<100$~ms.  In addition de Waele and Aarts quote an amplitude only for two initially antiparallel vortices; other initial orientations might yield different values for $B$.  On the other hand, we observe only a two-dimensional projection of each reconnection event, which would lead us to underestimate $B$, potentially furthering the discrepancy.  Clearly, resolving the source of this discrepancy warrants additional investigation.

The predicted scaling of $\alpha=0.5$ is derived from the assumption that the quantum of circulation $\kappa$ is the only relevant parameter over the length- and time-scales of interest.  This assumption is valid in the context of line-vortex methods, and holds approximately for the Gross-Pitaevskii equation at length scales large compared with the core diameter.  However, deviations from $\alpha=0.5$ might be obtained, at least conceptually, in two ways.  First, adapting arguments that were proposed for multiscaling solutions of the Euler equation related to intermittency in classical fluid turbulence \cite{sreenivasan88}, one might suppose the precise value of $\kappa$ is irrelevant to the dynamics of reconnection on the length scales we observe; then it should be possible to form a continuous family of solutions with differing values of $\alpha$.  However, it is difficult to imagine conditions under which $\kappa$ would be irrelevant in our experiments given that the observed velocity magnitudes seem closely related to $\kappa$ and to the distances involved, as expected.

\begin{figure}[b]
\begin{center}
\includegraphics[width=.48\textwidth]{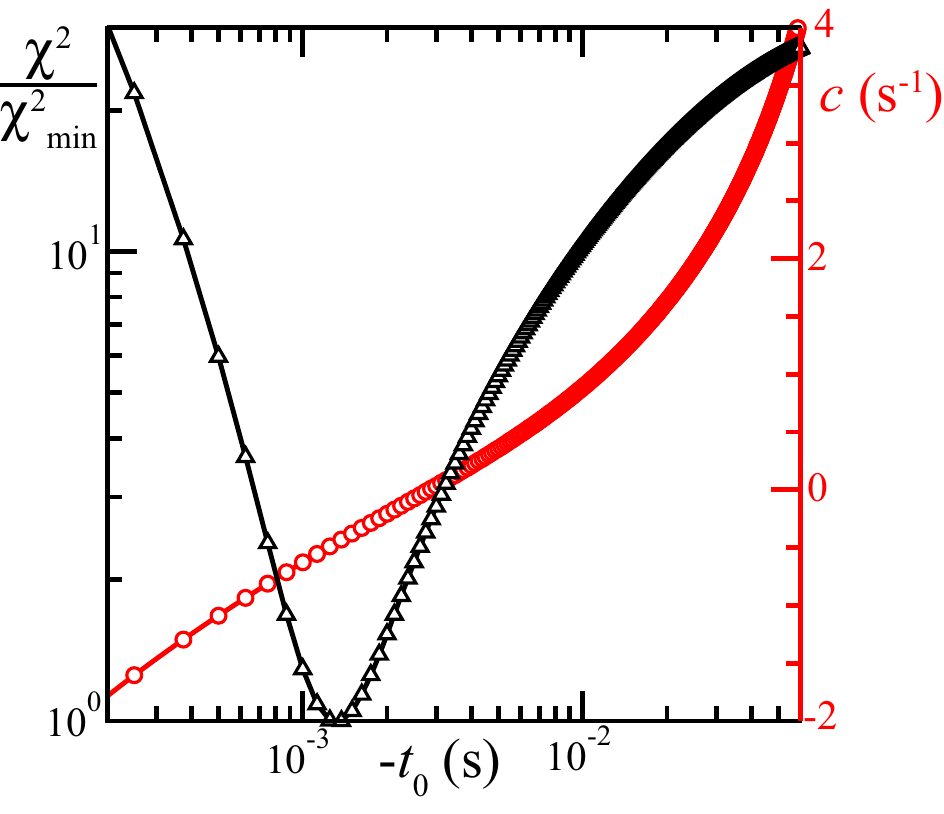}
\caption{Variation of $\chi^2(t_0)$ normalized by its minimum value $\chi^2_{\mathrm{min}}=0.73$ (black triangles) and the corresponding correction factor $c$ (red circles) as a function of the time origin $t_0$ for the event shown by the red squares in Fig.\ \ref{correction_events}.  We choose the parameters of the correction-factor expression \{$A$, $c$, $t_0$\} that minimize $\chi^2$ as defined by (\ref{chi2eqn}).}
\label{chi2_vs_t0}
\end{center}
\end{figure}

Second, if another parameter with units different from $\kappa$ were relevant then it is possible to rationalize reconnection dynamics with $\alpha \neq 0.5$.  If any parameter such as a vortex-core length-scale, core surface tension, typical intervortex spacing, local velocity gradients, or system size, were relevant to the reconnection dynamics, then we may construct putative solutions with variable values of $\alpha$.  For example, if a core surface tension $\gamma$ were relevant, we could contemplate an expression of the form
\begin{equation}
\delta(t) = B\vert t-t_0 \vert ^{\alpha} \kappa^{2-3\alpha} (\gamma/\rho) ^{2\alpha -1},
\label{altform}
\end{equation}
where we use the density $\rho$ to construct a kinematic surface tension $\gamma/\rho$.  Note that for $\alpha=0.5$ we recapture the predicted behavior (\ref{delta}).  The value of $\alpha$ for a particular reconnection event, in this interpretation, is either determined by the allowed values of $\alpha$ from the nonlinear equations of motion, or should that not be unique,  additionally by the initial and boundary data for each particular event.  Further deeper theoretical investigations are required to determine if such solutions could be realized under experimental conditions.

\subsection{Correction-Factor Expression}

\begin{figure}[t]
\begin{center}
\includegraphics[width=.48\textwidth]{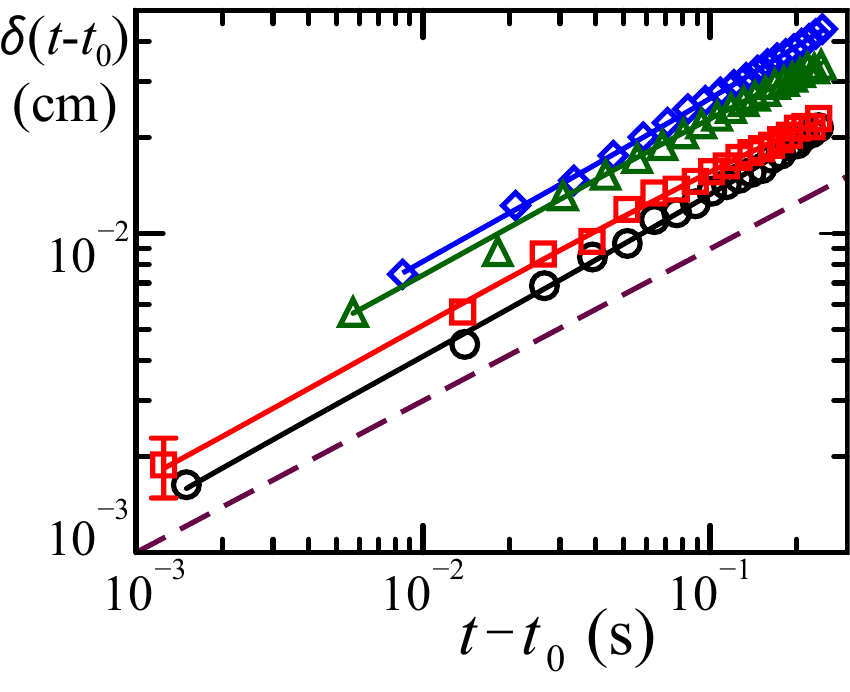}
\caption{Four forward events well fit by the correction-factor expression.  Symbols denote the separation $\delta(t)$ of pairs of particles on reconnecting vortices with an example error bar $\sigma=4$ $\mu$m while solid lines show fits to the correction-factor expression (\ref{fisherfit2}).  The predicted asymptotic form $\delta(t) =(\kappa\vert t-t_0\vert)^{1/2}$ is shown by the purple dashed line.}
\label{correction_events}
\end{center}
\end{figure}

\begin{figure}[t]
\begin{center}
\includegraphics[width=.48\textwidth]{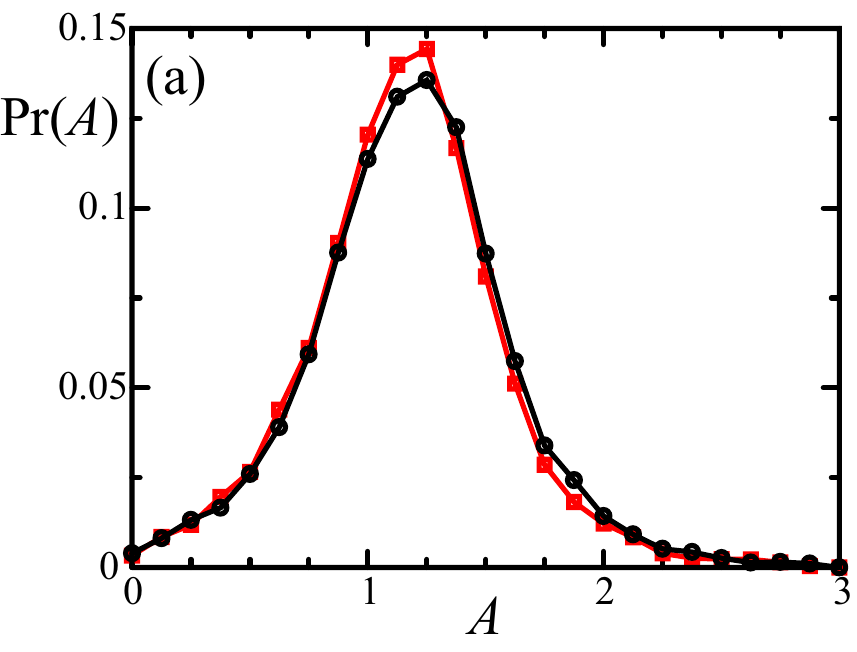}
\includegraphics[width=.48\textwidth]{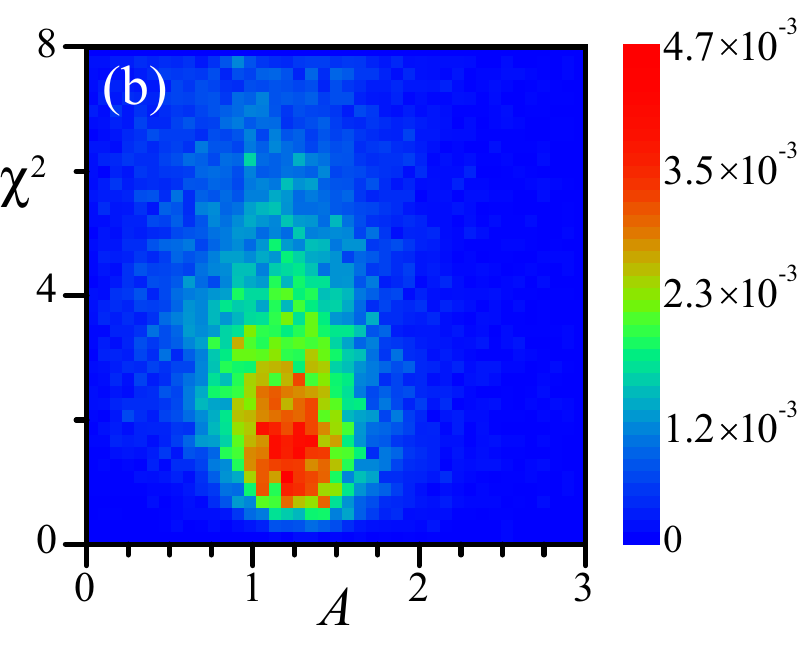}
\caption{(a) Normalized frequency distributions of the amplitude $A$ for 19,600 forward events (black circles) and 19,300 reverse events (red squares).  Both distributions are broad with a peak at $A=1.25$,  the means of the forward and reverse distributions of $A$ being 1.25 and 1.23, respectively.  (b) Two-dimensional contour diagram of $\chi^2$ versus $A$ for the forward events.  The peak near $A=1$ at low values of $\chi^2$ indicates that (\ref{fisherfit2}) describes optimally events with dynamics close to those predicted in (\ref{delta}).}
\label{Adists}
\end{center}
\end{figure}

\begin{figure}[t]
\begin{center}
\includegraphics[width=.48\textwidth]{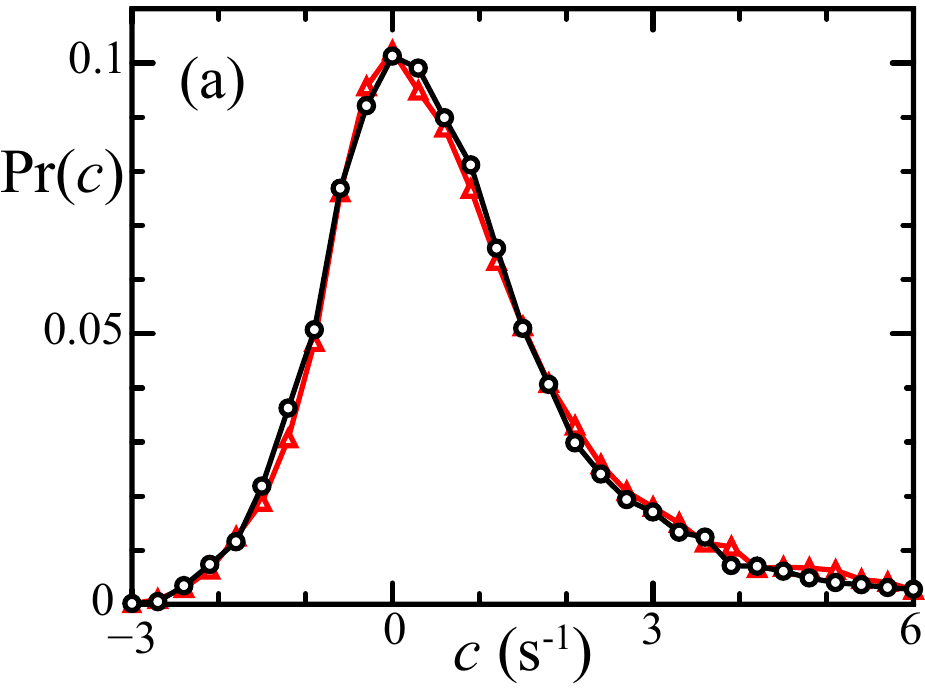}
\includegraphics[width=.48\textwidth]{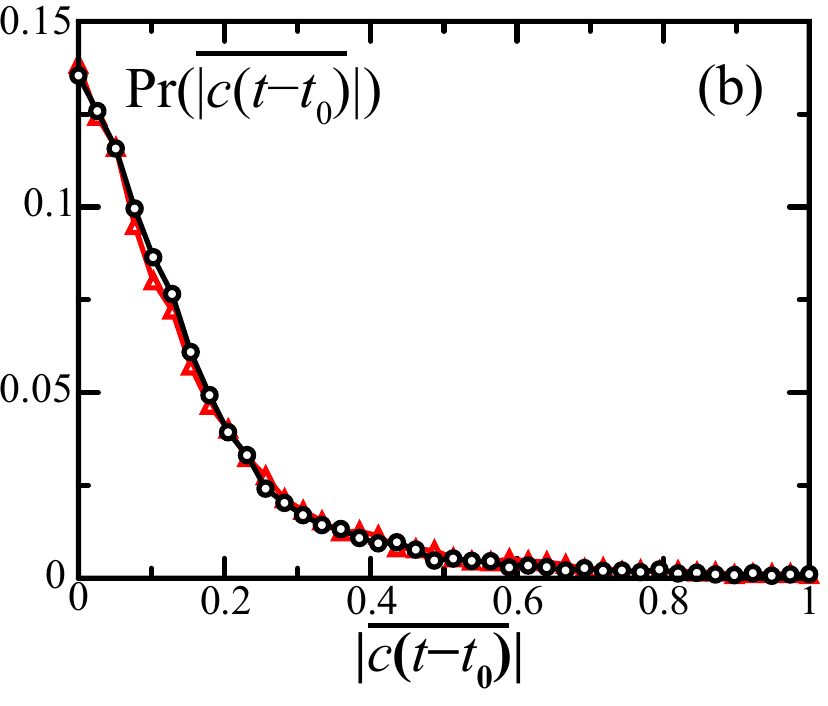}
\caption{(a) Normalized frequency distributions of the correction amplitude $c$ for 19,600 forward events (black circles) and 19,300 reverse events (red triangles).  (b) Normalized distributions of the magnitude of the correction factor $\overline{|c(t-t_0)|}$ time-averaged over 0.25 s for each trajectory for the forward (black circles) and reverse (red triangles) events.}
\label{Cdists}
\end{center}
\end{figure}

The dynamics of reconnection may alternatively be described by supplementing the predicted asymptotic scaling of (\ref{delta}) with a correction factor.  The simplest and natural expectation is the three-parameter form
\begin{equation}
\delta(t) \approx A\left (\kappa \vert t-t_0 \vert \right ) ^{1/2} \left (1+c \vert t-t_0 \vert \right ).
\label{fisherfit2}
\end{equation}
To test this expression we have performed a linear least-squares fit to determine $A$ and $c$ for 500 values of $t_0$ evenly spaced 125~$\mu$s apart (see Fig.\ \ref{chi2_vs_t0}); we then select the set of \{$A$, $c$, $t_0$\} that minimizes $\chi^2$ as defined in (\ref{chi2eqn}).

The measured separations $\delta(t)$ for four forward events are shown in Fig.\ \ref{correction_events}, along with the predicted scaling of (\ref{delta}) for comparison.  Solid line fits to the correction-factor expression in (\ref{fisherfit2}) describe the data well.  Distributions of the amplitude $A$, computed from the same fifty distinct experimental heat pulses used to form Figs. \ref{alpha_dists} and \ref{B_dists}, are shown in Fig.\ \ref{Adists}(a).  As before, we require $\chi^2<4$.  For both the forward and reverse events, the distributions of $A$ are peaked near unity, in accord with the dimensional arguments; however as in Fig.\ \ref{B_dists}, the values typically exceed unity (so being about three times greater than found in the special case studied numerically by de Waele and Aarts \cite{deWaele94}).  Events with $A$ near unity typically have lower values of $\chi^2$, as shown in Fig.\ \ref{Adists}(b), again supporting the inferences based on dimensional analysis.

The distributions of the correction amplitude $c$ for the forward and reverse events are shown in Fig.\ \ref{Cdists}(a).  The distributions are peaked at $c=0$, indicating that many events follow rather closely the simple scaling of (\ref{delta}).  However, for both the forward and reverse events, the distributions are broad compared to their mean values of 0.63 and 0.71 s$^{-1}$, respectively, signifying strong event to event variation.  The correction factor also varies systematically with the fitted value of the time origin $t_0$, as illustrated in Fig.\ \ref{chi2_vs_t0} and easily understood algebraically.  Thus, uncertainties in the estimation of $t_0$ will induce corresponding changes in estimates for $c$.

It is useful to consider the magnitude of the correction term, since larger magnitudes imply a greater departure from the asymptotic form (\ref{delta}).  This departure can be quantified by $\overline{|c(t-t_0)|}$, where the overbar implies a time-average over the duration that we use to fit the data, namely $0<\vert t-t_0 \vert <0.25$~s.  The ensemble mean value of $\overline{|c(t-t_0)|}$ for both the forward and reverse events is $\langle \overline{|c(t-t_0)|} \rangle =0.15$, but it ranges from $10^{-6}$ to values greater than unity.  This implies that while the deviations vary from event to event, they typically amount to less than $\pm 20$~\% as evidenced in Fig.\ \ref{Cdists}(b).

The need to supplement the simple form (\ref{delta}), derived by dimensional analysis, with a correction factor stems from several potential sources.  Independent of a specific origin, one must always expect a subdominant term in asymptotic power-law scaling forms like (\ref{delta}) which describe behavior from a micro- or mesoscopic domain up to some appropriate infrared cutoff at long times or large length-scales.  A functional form including a correction factor along with the dominant power-law allows for a crossover between scales.  In our case, one certainly should expect deviations from (\ref{delta}) on length scales comparable to the typical intervortex spacing of 0.1 to 1.0~mm.  Indeed, if one introduces a correction length scale, say $l$, by rewriting the correction term as
\begin{equation}
c\vert t-t_0 \vert \equiv \pm \kappa \vert t-t_0 \vert /l^2,
\label{correction_length}
\end{equation}
for forward and reverse trajectories, one finds that the corresponding forward and reverse mean values of $c$ correspond to $l=0.40$~mm and 0.38~mm, respectively.  Thus the dominant correction may well represent the influence of neighboring vortices and their ability to distort the observed trajectories.

In addition, however, other spatial and temporal aspects of the local environment may also significantly affect the dynamics of reconnection, beyond the leading behavior accounted for by dimensional analysis.  Local velocity gradients or other initial and boundary conditions could all produce deviations from pure square-root scaling, thereby necessitating a nonzero correction factor.  More intriguing theoretically, however, and challenging experimentally, is the possibility of nonanalytic correction terms such as $c_{\theta} \vert t-t_0 \vert ^{\theta}$ with $\theta$ nonintegral.  The presence of such terms with nontrivial values of $\theta \simeq 0.5$ is well established in the study of critical phenomena in ferromagnets, superfluids, at gas-liquid transitions, etc.: see, e.g., \cite{wegner72, chen82, guida97}.

\subsection{Comparative Assessment}
\label{Comparison}

\begin{figure}[b]
\begin{center}
\includegraphics[width=.48\textwidth]{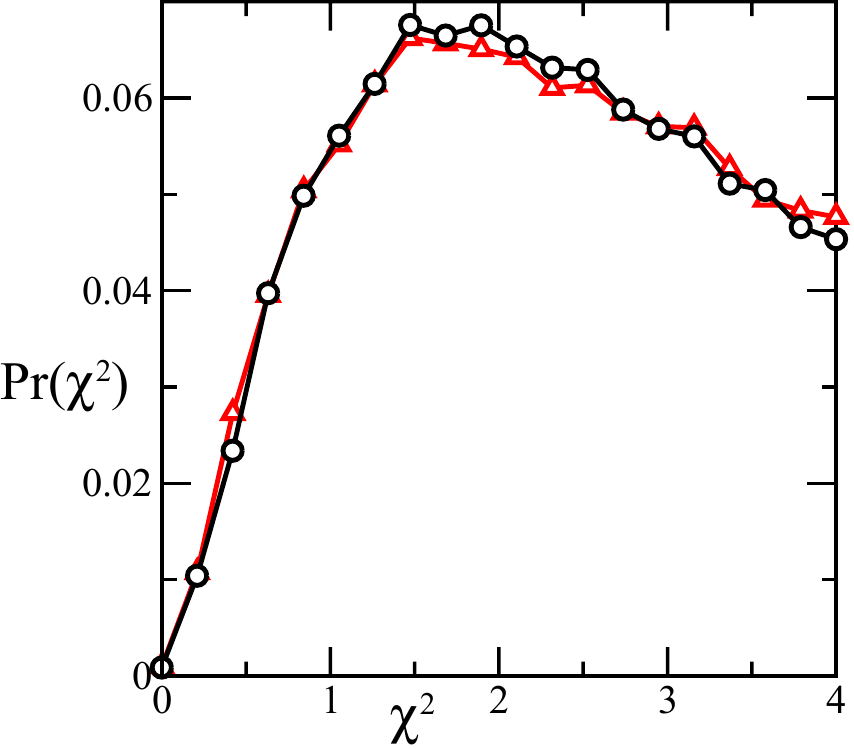}
\caption{Normalized frequency distributions of $\chi^2$ from both the forward and reverse events fit to the arbitrary power-law expression (black circles) and the correction-factor expression (red triangles).}
\label{chidists}
\end{center}
\end{figure}

The two expressions (\ref{alphafit2}) and (\ref{fisherfit2}) are both modified versions of the asymptotic dynamics which suggest somewhat distinct theoretical interpretations.  In both cases, the majority of the fitted events exhibit behavior very similar to the dimensional predictions (i.e., $\alpha=0.5$ or $c \simeq 0$); although, both also show strong variations from event to event.  One might hope to distinguish the quality of the two fits by comparing the observed distributions of $\chi^2$, the overall deviation in the fits.  In fact and unsurprisingly, the two distributions shown in Fig.\ \ref{chidists} are rather similar and no firm basis for making any distinctions emerges.

On balance at this point we favor the correction-factor expression as best embodying our experimental data for quantized vortex reconnection.  Our typical observations of the dynamics show only relatively slight deviations from those predicted by dimensional analysis.  It therefore appears that the dominant parameter is indeed the quantum of circulation $\kappa$, which sets the leading scaling exponent of $\alpha=0.5$.  We expect the deviations from the corresponding asymptotic form in our experimental range to be caused by the local environment and the initial and boundary conditions of the event, as opposed to other parameters such as a surface tension of the vortex cores.  Indeed, such parameters would likely vary with temperature and we have not observed any correlations between our fit parameters and the temperature of the system over the range 1.70~K$<T<2.05$~K.  Experiments and numerical simulations that control the local environment (velocity gradients, neighboring vortices, strains, etc.) as well as the initial and boundary conditions (configuration of the vortices, initial velocities and curvatures, etc.) could directly test this hypothesis and are clearly desirable.

\begin{figure*}[t]
\begin{center}
\includegraphics[height=0.31\textwidth]{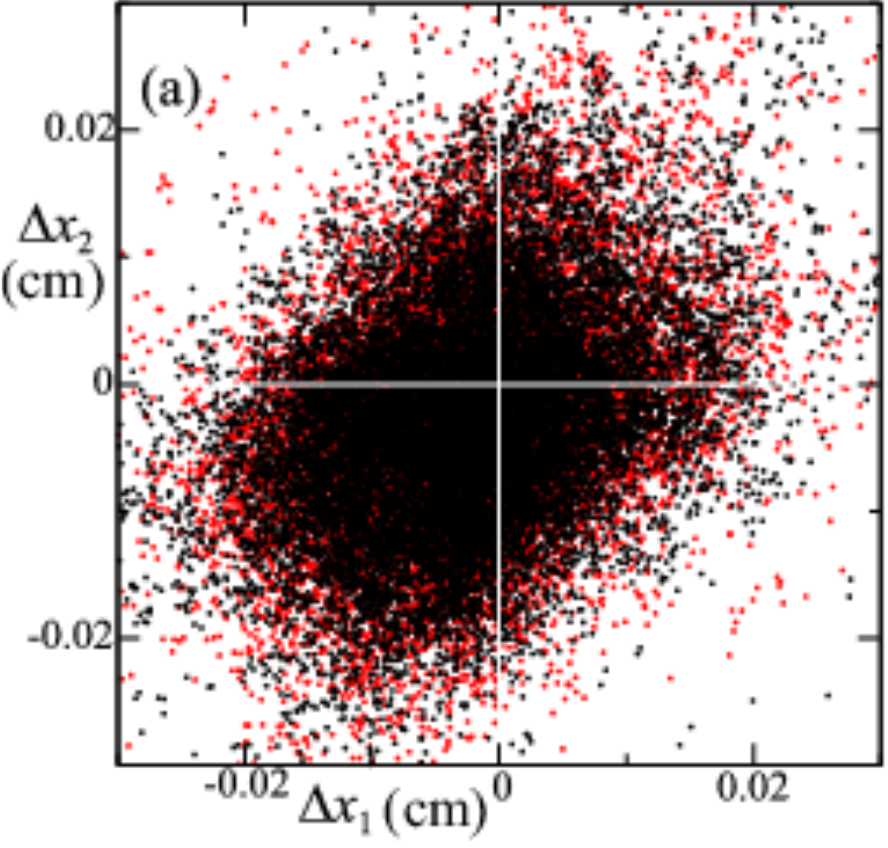}
\includegraphics[height=0.31\textwidth]{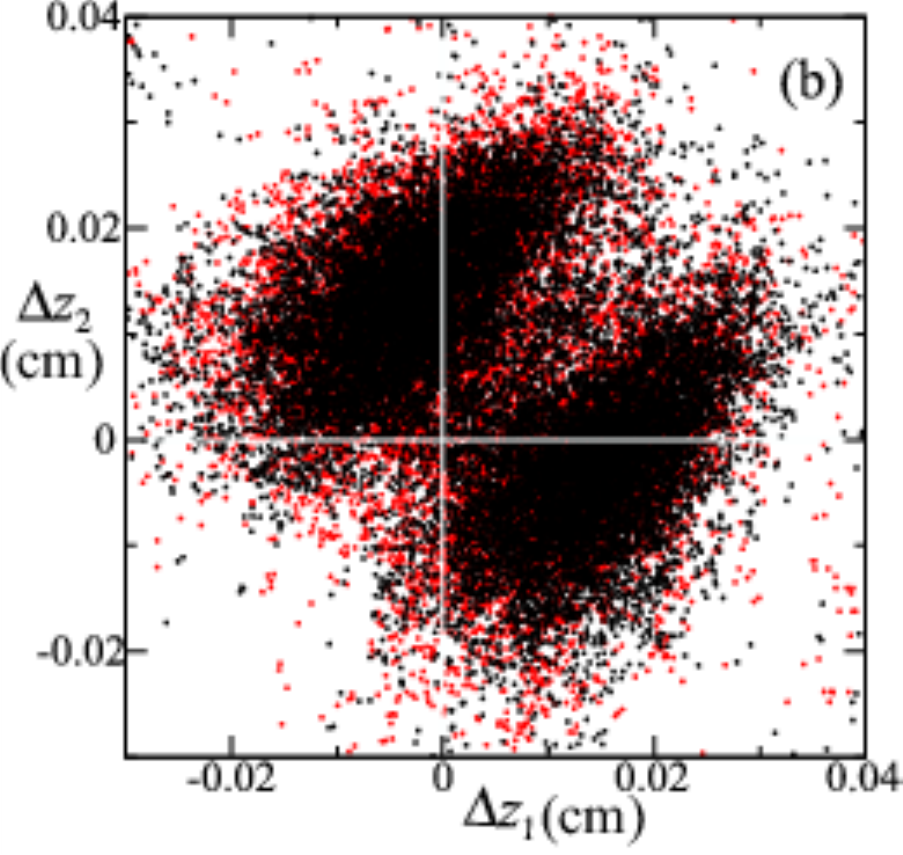}
\includegraphics[height=0.31\textwidth]{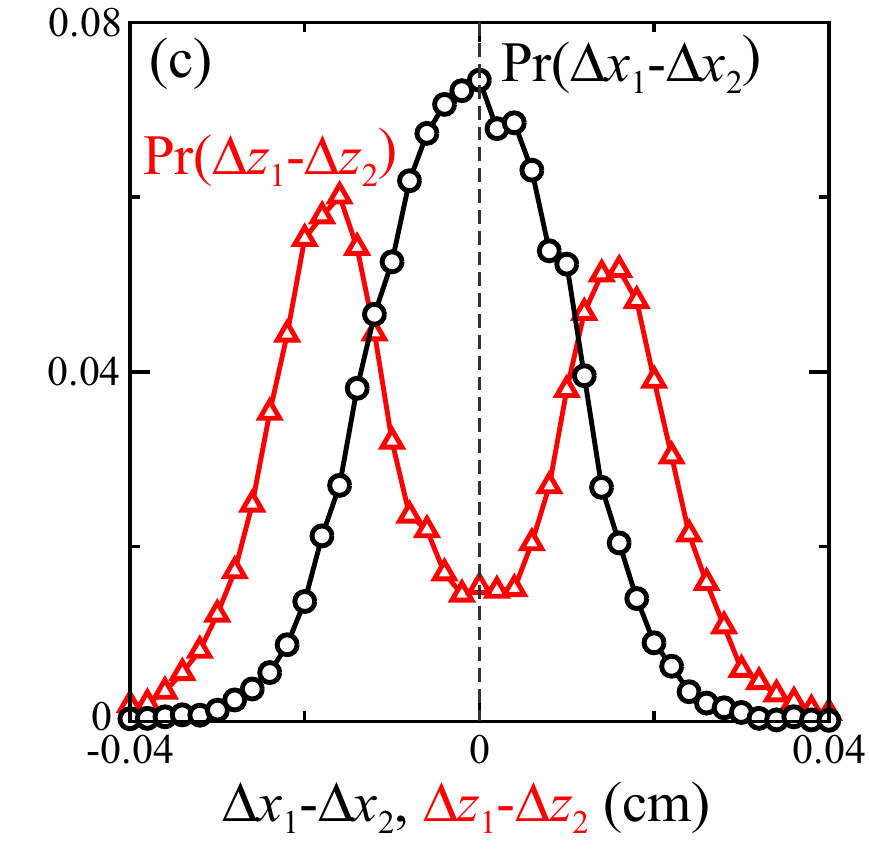}
\caption{Scatter plots of (a) the $\hat{x}$-component and (b) the $\hat{z}$-component of the displacement vector $\Delta \mathbf{r}_i=\mathbf{r}_i(t_0+0.25$~s)$-\mathbf{r}_i(t_0)$ for all forward (black) and reverse (red) event pairs (labeled as $i=1$ and $j=2$ for all pairs).  (c) Normalized frequency distributions of the differences in $\hat{x}$-components (black circles) and the $\hat{z}$-components (red triangles) of the displacement vectors for each pair of particles on reconnecting vortices.  The central peak at $\Delta x_1-\Delta x_2=0$ signifies a strong correlation of the particle trajectories in the $x$-direction while the minimum at $\Delta z_1-\Delta z_2=0$ implies a strong anti-correlation in the $z$-direction associated with the $z$-directed thermal counterflow.}
\label{delta_scatter}
\end{center}
\end{figure*}

\section{Time-Reversibility and Anisotropy}
\label{Time-Reversibility}

The Gross-Pitaevskii equation as used for quantized vortex reconnection \cite{koplik93, nazarenko03} is fully symmetric under time reversal; thus solutions of the equation may also be time-reversible symmetric.  However, many previous theoretical works have concluded that reconnection dissipates energy by emitting acoustic and Kelvin waves \cite{leadbeater01, vinen01, ogawa02, vinen05} that may be absorbed by the boundaries, and so would break locally the time-reversibility.  We have compared the pre- and post-reconnection dynamics by separately fitting forward and reverse events.  All of the distributions of the fit parameters (see Figs.\ \ref{alpha_dists}, \ref{B_dists}, \ref{Adists} and \ref{Cdists}) for the forward and reverse events show striking similarity, as would be expected if the dynamics were statistically time-reversible.  While it is clear, however, that some energy is dissipated overall in our experiments, as evidenced by the decay of the turbulent state, it is not evidenced in the statistics of individual events.

In addition to the fit parameters, we may also investigate the total displacement of the vortices before and after events.  We define the displacement vector of particle $i$ as
\begin{equation}
\Delta \mathbf{r}_i=\mathbf{r}_i(t_0+0.25 \; \mathrm{s})-\mathbf{r}_i(t_0)=\hat{x}\Delta x_i +\hat{z} \Delta z_i.
\end{equation}
Figs.\ \ref{delta_scatter}(a) and (b) show the $\hat{x}$- and $\hat{z}$-components of the displacement vectors for all of the particles identified with forward (black) and reverse (red) reconnection events.  Indeed, the forward and reverse displacement vectors also show striking similarities.  The displacement vectors appear weakly correlated in the $x$-direction and anti-correlated in the $z$-direction.  This anisotropy is clearly exhibited in Fig.\ \ref{delta_scatter}(c), which shows the difference in the displacement vectors of the pairs of particles on reconnecting vortices.  The forward and reverse events are found to be equally affected by the these anisotropic effects.

We believe the anisotropy arises from the polarizing effect of the $z$-directed, initiating thermal counterflow.  A possible interpretation of the anti-correlation in the $z$-direction is that the vortices are typically aligned or anti-aligned with the direction of the counterflow.  Previous studies have also observed \cite{wang87} or argued for \cite{gordeev05, barenghi06, barenghi07} the presence of anisotropy in counterflow turbulence; although, subject to interpretation, it is not clear that we agree on the polarization of the anisotropy.

Overall the close statistical similarity of the forward and reverse events suggests an effective equilibrium has been established in quantum turbulence on the time scales ($\leq 0.25$~s) we have investigated.  We have only limited statistics to directly compare the dynamics before and after a given event, the details of which will be reserved for a future publication.  However, they are sufficient to show that individual events are not time-reversal invariant.  An important future direction for both experiments and numerical simulations entails understanding the interplay between irreversible individual events and the reversible statistics of quantized vortex reconnection.

\section{Effects of Reconnection on Quantum Turbulence}
\label{QT}

Reconnection has long been considered to play an important dissipative role in quantum turbulence.  Vinen \cite{vinen57a, vinen57b, vinen57c} described how the balance of reconnection and mutual friction leads to saturated vortex line lengths in counterflow turbulence, which is analogous to the saturation of dynamo action \cite{vainshtein92} or the magnetorotational instability \cite{balbus98} produced by magnetic reconnection in astrophysical plasmas.  A great deal of recent research, though, has focussed on the behavior of quantum turbulence on length scales sufficiently large that the interactions of individual quantized vortices may be neglected.  These previous works concluded that on such length scales quantum turbulence shares many characteristics with classical turbulence \cite{smith93, barenghi97, nore97, maurer98, barenghi99, stalp99, vinen00, skrbek00, vinen02, barenghi02a, barenghi02b, skrbek03, kobayashi05, kobayashi06, kobayashi07, l'vov07, chagovets07, morris08}.  However, the strength of the evidence has been questioned \cite{procaccia08}.

The assumptions used to argue for the classical nature of quantum turbulence break down on length scales smaller than the typical intervortex spacing, which is what we have probed with our measurements.  It is clear from the online movies (see \cite{paoletti08a}) that our turbulent states in superfluid $^4$He differ drastically from those observed in classical fluids as a result of the topological interactions of the quantized vortices.  Specifically, reconnection produces anomalously large velocities in highly-localized areas, which are not diffusively smoothed by viscosity.  If length-scales are evolving asymptotically as $R(t)=\tilde{A}\vert \kappa (t-t_0) \vert ^{1/2}$, then we expect the velocities to scale as
\begin{equation}
v(t)=\frac{\tilde{A}}{2} \sqrt{\frac{\kappa}{\vert t-t_0 \vert}},
\label{vscaling}
\end{equation}
which far exceed typical fluid velocities when $t\rightarrow t_0$.  Note, however, that we expect such velocities to be cut-off by the speed of first sound.

\begin{figure}[t]
\begin{center}
\includegraphics[width=.48\textwidth]{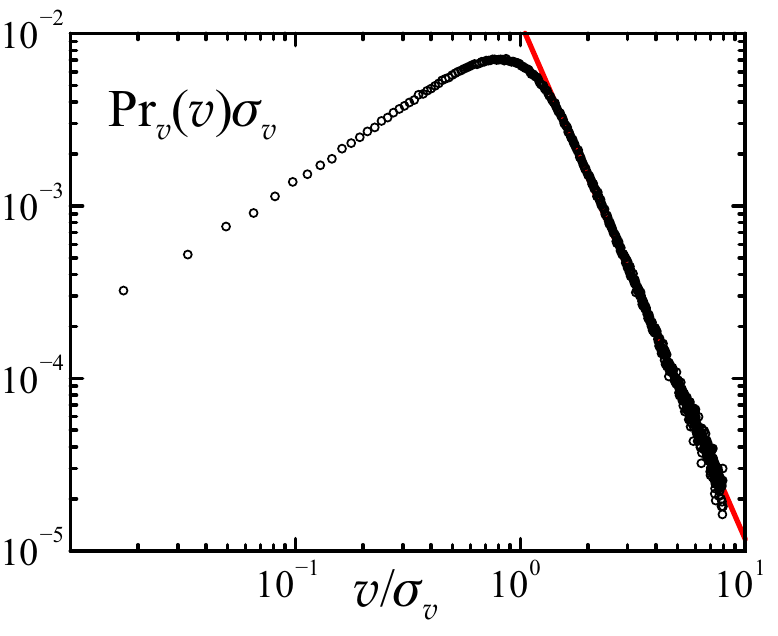}
\caption{Frequency distribution function of tracer particle velocities resulting from the decay of a turbulent state produced by a 0.17 W/cm$^2$ heat flux at a temperature $T=1.90$~K.  The distribution is scaled by the standard deviation $\sigma_v=0.08$ cm/s.  The straight line (in red) is a fit to Pr$_v(v)=av^{-3}$ for $v/\sigma_v > 1.5$.}
\label{vpdf}
\end{center}
\end{figure}

To model the velocity statistics of a turbulent state in superfluid $^4$He characterized by many reconnection events we may use the transformation
\begin{equation}
\rm{Pr}\it_v(v)=|dt/dv|\rm{Pr}\it _t[t(v)],
\end{equation}
where Pr$_v(v)dv$ is the probability of observing a velocity between $v$ and $v+dv$ at any time while Pr$_t(t)dt$ is the uniform
probability of taking a measurement at a time between $t$ and $t+dt$.  Accepting the relation (\ref{vscaling}), we predict for large $v$ (small $\vert t-t_0 \vert$) the behavior \cite{paoletti08a}
\begin{equation}
\rm{Pr}\it_v(v)\propto|dt/dv| \propto |v|^{\rm{-3}}.
\label{vfit}
\end{equation}

A probability distribution function of the velocities derived from \textit{all} tracer trajectories (not only those identified as marking reconnection) is shown in Fig.\ \ref{vpdf}.  The velocities are computed by forward differences of the particle trajectories.  A single-parameter fit of the form Pr$_v(v)=av^{-3}$, where $v=(v_x^2+v_z^2)^{1/2}$, is shown as a solid line for comparison.  Evidently, the simple arguments used to derive (\ref{vfit}) are able to predict the tails of the velocity distributions in reconnection-dominated quantum turbulence.

The velocity statistics of classical turbulence \cite{frisch95} in both experiments \cite{noullez97} and numerical simulations \cite{vincent91, gotoh02} are nearly Gaussian over several orders of magnitude in probability.  Such statistics are in stark contrast to the power-law statistics found in Fig.\ \ref{vpdf}.  We attribute this distinction to the topological interactions of the quantized vortices, which do not exist in classical turbulence where the velocity field is diffusively smoothed by viscosity.  It is important to note that the normal fluid present in our experiments is relatively quiescent and potential future directions could include examining the velocity statistics for the case where both the normal fluid and superfluid are turbulent, since the two fluids couple through friction acting on the quantized vortices \cite{vinen57a, vinen57b, vinen57c}.

\section{Implications for Classical Turbulence}
\label{CT}

The scaling properties of velocity correlations in classical turbulence have received a great deal of attention.  In particular, much debate has addressed the values of the exponents $\zeta_n$ of the longitudinal structure functions $\langle \Delta u_r^n \rangle \sim r^{\zeta_n}$, where $\Delta u_r \equiv u(\mathbf{x}+\mathbf{r})-u(\mathbf{x})$ for a single velocity component $u$ parallel to $\mathbf{r}$ \cite{kolmogorov41a, kolmogorov62, benzi84, anselmet84, meneveau87, andrews89, kida91, she91, vincent91, stolovitzky93, benzi93a, benzi93b, she94, barenblatt95, frisch95, arneodo96, boratav97, sreenivasan97, lewis99, chevillard05}.  In analogy to the predictions of $\alpha=0.5$ for quantized vortex reconnection, Kolmogorov used dimensional arguments to predict $\zeta_n=n/3$ \cite{kolmogorov41a}.  His theory pivots on the assumption that the dissipation per unit mass $\epsilon$ is the only relevant parameter in the observed correlations and spectra.  However, experimental observations report values of $\zeta_n$ that deviate slightly from Kolmogorov scaling \cite{anselmet84, benzi93a, benzi93b, arneodo96, lewis99}.  Typically, arbitrary values of the exponents $\zeta_n$ are fit to the data \cite{vincent91, anselmet84, kolmogorov62, benzi84, meneveau87, andrews89, kida91, she91, stolovitzky93, benzi93a, benzi93b, she94, frisch95, boratav97, sreenivasan97, lewis99, chevillard05}, in analogy to our arbitrary power-law expression in (\ref{alphafit2}).  It should be noted, however, that models with variable $\zeta_n$ presuppose that arbitrary exponents are allowed, based either on the irrelevance of $\epsilon$, or on dimensional grounds, by the admission of other relevant quantities that yield new power law forms [as illustrated in (\ref{altform})].

We argue, though, that another option is available -- that of correction-factors representing subdominant scalings. Similar connections between critical phenomena and turbulence have been explored previously \cite{fineburg93, barenblatt95}.  Here we specifically suggest that individual events in classical turbulence might be modeled both by a dominant Kolmogorov term and a correction factor arising from various causes including at least the local neighborhood conditions and finite-size effects.  The basic Kolmogorov scaling derives from the fact that with vanishing viscosity one obtains $\langle \Delta u_r^3 \rangle = - \frac{4}{5} \epsilon r$.  If one interprets these statistics as stemming from many individual \lq \lq Kolmogorov events" with $\Delta u_r \sim \delta /  (t-t_0)$, then substituting $r \rightarrow \delta$ above yields, at least on a dimensional basis, $\delta \sim \epsilon^{1/2} (t-t_0)^{3/2}$, which is also known as Richardson scaling \cite{richardson26}.  This can be obtained directly from a dimensional argument if $\delta$ depends only on $\epsilon$ and time.  Note the units of $\epsilon$ are $\mbox{m}^2/\mbox{s}^3$.  If one then extends this model to a correction-factor expression, similar to (\ref{fisherfit2}), with an appropriately chosen correction, one might obtain behavior that would be difficult to distinguish from fluctuating power laws, though with a rather different interpretation.

\section{Conclusions}
\label{Conclusions}

In conclusion, we have observed the dynamics of individual reconnection events in superfluid $^4$He and their effects on decaying quantum turbulence.  Although we observe significant deviations that vary from event to event from the mean behavior, the typical dynamics are close to those predicted by dimensional arguments.  We regard this as our major finding.  The deviations may be accounted for in two separate ways: (a) by supposing the scaling exponent of the dynamics can fluctuate as in (\ref{alphafit2}), or (b) by recognizing that the predicted power laws must be supplemented by a correction factor such as in (\ref{fisherfit2}).  The two three-parameter expressions describe the data almost equally well from a $\chi^2$ perspective, but suggest distinct physical interpretations.  Thus, we observe that a variable scaling exponent should result from either a lack of importance of the precise value of the quantum of circulation $\kappa$ or from the competing relevance of another physical quantity of distinct dimensions (such as a length-scale, surface tension, etc.).

On the other hand and more naturally, we interpret the correction factor as arising from initial conditions and boundary effects, such as the vorticity distribution and intervortex spacing, and from properties of the local environment at reconnection, such as velocity gradients, pressure gradients, and thermal fluctuations.  Since the dynamics appear to be well characterized by the predictions that assume that the only relevant physical parameter is the quantum of circulation $\kappa$, we believe our data indicate that the environment, as opposed to other parameters, is most likely the origin of the observed deviations.  Further investigations experimentally and theoretically could focus on: (\textit{i}) determining if each reconnection event is time-reversal symmetric, (\textit{ii}) considering alternate forms of the correction term, such as $c_{\theta} \vert t-t_0 \vert ^{\theta}$ with $\theta \ne 1$, and (\textit{iii}) systematically changing the initial and boundary conditions as well as the local environment near reconnection and investigating the resulting deviations from the dimensionally predicted asymptotic form.

We thank Makoto Tsubota, Carlo Barenghi, Joseph Vinen, Nigel Goldenfeld, Christopher Lobb, Marc Swisdak, and James Drake for helpful discussions.  This work was supported by NSF DMR-0606252, NSF PHY05-51164 and the Center for Nanophysics and Advanced Materials at the University of Maryland.

% \appendix

% \section{}
% \label{}

\ed
\begin{thebibliography}{101}
\expandafter\ifx\csname natexlab\endcsname\relax\def\natexlab#1{#1}\fi
\expandafter\ifx\csname bibnamefont\endcsname\relax
  \def\bibnamefont#1{#1}\fi
\expandafter\ifx\csname bibfnamefont\endcsname\relax
  \def\bibfnamefont#1{#1}\fi
\expandafter\ifx\csname citenamefont\endcsname\relax
  \def\citenamefont#1{#1}\fi
\expandafter\ifx\csname url\endcsname\relax
  \def\url#1{\texttt{#1}}\fi
\expandafter\ifx\csname urlprefix\endcsname\relax\def\urlprefix{URL }\fi
\providecommand{\bibinfo}[2]{#2}
\providecommand{\eprint}[2][]{\url{#2}}

\bibitem[{\citenamefont{Donnelly}(1991)}]{donnelly91}
\bibinfo{author}{\bibfnamefont{R.~J.} \bibnamefont{Donnelly}},
  \emph{\bibinfo{title}{Quantized Vortices in Helium II}}
  (\bibinfo{publisher}{Cambridge Univ. Press}, \bibinfo{address}{Cambridge,
  UK}, \bibinfo{year}{1991}).

\bibitem[{\citenamefont{{Chuang} et~al.}(1991)\citenamefont{{Chuang}, {Yurke},
  {Durrer}, and {Turok}}}]{chuang91}
\bibinfo{author}{\bibfnamefont{I.}~\bibnamefont{{Chuang}}},
  \bibinfo{author}{\bibfnamefont{B.}~\bibnamefont{{Yurke}}},
  \bibinfo{author}{\bibfnamefont{R.}~\bibnamefont{{Durrer}}}, \bibnamefont{and}
  \bibinfo{author}{\bibfnamefont{N.}~\bibnamefont{{Turok}}},
  \bibinfo{journal}{Science} \textbf{\bibinfo{volume}{251}},
  \bibinfo{pages}{1336} (\bibinfo{year}{1991}).

\bibitem[{\citenamefont{{Blatter} et~al.}(1994)\citenamefont{{Blatter},
  {Feigel'Man}, {Geshkenbein}, {Larkin}, and {Vinokur}}}]{blatter94}
\bibinfo{author}{\bibfnamefont{G.}~\bibnamefont{{Blatter}}},
  \bibinfo{author}{\bibfnamefont{M.~V.} \bibnamefont{{Feigel'Man}}},
  \bibinfo{author}{\bibfnamefont{V.~B.} \bibnamefont{{Geshkenbein}}},
  \bibinfo{author}{\bibfnamefont{A.~I.} \bibnamefont{{Larkin}}},
  \bibnamefont{and} \bibinfo{author}{\bibfnamefont{V.~M.}
  \bibnamefont{{Vinokur}}}, \bibinfo{journal}{Rev. Mod. Phys.}
  \textbf{\bibinfo{volume}{66}}, \bibinfo{pages}{1125} (\bibinfo{year}{1994}).

\bibitem[{\citenamefont{Priest and Forbes}(2000)}]{priest00}
\bibinfo{author}{\bibfnamefont{E.}~\bibnamefont{Priest}} \bibnamefont{and}
  \bibinfo{author}{\bibfnamefont{T.}~\bibnamefont{Forbes}},
  \emph{\bibinfo{title}{Magnetic Reconnection: MHD Theory and Applications}}
  (\bibinfo{publisher}{Cambridge Univ. Press}, \bibinfo{address}{Cambridge,
  UK}, \bibinfo{year}{2000}).

\bibitem[{\citenamefont{{Lin} and {Hudson}}(1971)}]{lin71}
\bibinfo{author}{\bibfnamefont{R.~P.} \bibnamefont{{Lin}}} \bibnamefont{and}
  \bibinfo{author}{\bibfnamefont{H.~S.} \bibnamefont{{Hudson}}},
  \bibinfo{journal}{Sol. Phys.} \textbf{\bibinfo{volume}{17}},
  \bibinfo{pages}{412} (\bibinfo{year}{1971}).

\bibitem[{\citenamefont{{Lin} et~al.}(2003)\citenamefont{{Lin}, {Krucker},
  {Hurford}, {Smith}, {Hudson}, {Holman}, {Schwartz}, {Dennis}, {Share},
  {Murphy} et~al.}}]{lin03}
\bibinfo{author}{\bibfnamefont{R.~P.} \bibnamefont{{Lin}}},
  \bibinfo{author}{\bibfnamefont{S.}~\bibnamefont{{Krucker}}},
  \bibinfo{author}{\bibfnamefont{G.~J.} \bibnamefont{{Hurford}}},
  \bibinfo{author}{\bibfnamefont{D.~M.} \bibnamefont{{Smith}}},
  \bibinfo{author}{\bibfnamefont{H.~S.} \bibnamefont{{Hudson}}},
  \bibinfo{author}{\bibfnamefont{G.~D.} \bibnamefont{{Holman}}},
  \bibinfo{author}{\bibfnamefont{R.~A.} \bibnamefont{{Schwartz}}},
  \bibinfo{author}{\bibfnamefont{B.~R.} \bibnamefont{{Dennis}}},
  \bibinfo{author}{\bibfnamefont{G.~H.} \bibnamefont{{Share}}},
  \bibinfo{author}{\bibfnamefont{R.~J.} \bibnamefont{{Murphy}}},
  \bibnamefont{et~al.}, \bibinfo{journal}{Astrophys. J.}
  \textbf{\bibinfo{volume}{595}}, \bibinfo{pages}{L69} (\bibinfo{year}{2003}).

\bibitem[{\citenamefont{{Terasawa} and {Nishida}}(1976)}]{terasawa76}
\bibinfo{author}{\bibfnamefont{T.}~\bibnamefont{{Terasawa}}} \bibnamefont{and}
  \bibinfo{author}{\bibfnamefont{A.}~\bibnamefont{{Nishida}}},
  \bibinfo{journal}{Planet. Space Sci.} \textbf{\bibinfo{volume}{24}},
  \bibinfo{pages}{855} (\bibinfo{year}{1976}).

\bibitem[{\citenamefont{{Baker} and {Stone}}(1976)}]{baker76}
\bibinfo{author}{\bibfnamefont{D.~N.} \bibnamefont{{Baker}}} \bibnamefont{and}
  \bibinfo{author}{\bibfnamefont{E.~C.} \bibnamefont{{Stone}}},
  \bibinfo{journal}{Geophys. Res. Lett} \textbf{\bibinfo{volume}{3}},
  \bibinfo{pages}{557} (\bibinfo{year}{1976}).

\bibitem[{\citenamefont{Savrukhin}(2001)}]{savrukhin01}
\bibinfo{author}{\bibfnamefont{P.~V.} \bibnamefont{Savrukhin}},
  \bibinfo{journal}{Phys. Rev. Lett.} \textbf{\bibinfo{volume}{86}},
  \bibinfo{pages}{3036} (\bibinfo{year}{2001}).

\bibitem[{\citenamefont{{{\O}ieroset} et~al.}(2002)\citenamefont{{{\O}ieroset},
  {Lin}, {Phan}, {Larson}, and {Bale}}}]{oieroset02}
\bibinfo{author}{\bibfnamefont{M.}~\bibnamefont{{{\O}ieroset}}},
  \bibinfo{author}{\bibfnamefont{R.~P.} \bibnamefont{{Lin}}},
  \bibinfo{author}{\bibfnamefont{T.~D.} \bibnamefont{{Phan}}},
  \bibinfo{author}{\bibfnamefont{D.~E.} \bibnamefont{{Larson}}},
  \bibnamefont{and} \bibinfo{author}{\bibfnamefont{S.~D.}
  \bibnamefont{{Bale}}}, \bibinfo{journal}{Phys. Rev. Lett.}
  \textbf{\bibinfo{volume}{89}}, \bibinfo{pages}{195001}
  (\bibinfo{year}{2002}).

\bibitem[{\citenamefont{{Dmitruk} et~al.}(2003)\citenamefont{{Dmitruk},
  {Matthaeus}, {Seenu}, and {Brown}}}]{dmitruk03}
\bibinfo{author}{\bibfnamefont{P.}~\bibnamefont{{Dmitruk}}},
  \bibinfo{author}{\bibfnamefont{W.~H.} \bibnamefont{{Matthaeus}}},
  \bibinfo{author}{\bibfnamefont{N.}~\bibnamefont{{Seenu}}}, \bibnamefont{and}
  \bibinfo{author}{\bibfnamefont{M.~R.} \bibnamefont{{Brown}}},
  \bibinfo{journal}{Astrophys. J.} \textbf{\bibinfo{volume}{597}},
  \bibinfo{pages}{L81} (\bibinfo{year}{2003}).

\bibitem[{\citenamefont{{Holman} et~al.}(2003)\citenamefont{{Holman}, {Sui},
  {Schwartz}, and {Emslie}}}]{holman03}
\bibinfo{author}{\bibfnamefont{G.~D.} \bibnamefont{{Holman}}},
  \bibinfo{author}{\bibfnamefont{L.}~\bibnamefont{{Sui}}},
  \bibinfo{author}{\bibfnamefont{R.~A.} \bibnamefont{{Schwartz}}},
  \bibnamefont{and} \bibinfo{author}{\bibfnamefont{A.~G.}
  \bibnamefont{{Emslie}}}, \bibinfo{journal}{Astrophys. J.}
  \textbf{\bibinfo{volume}{595}}, \bibinfo{pages}{L97} (\bibinfo{year}{2003}).

\bibitem[{\citenamefont{{Drake} et~al.}(2005)\citenamefont{{Drake}, {Shay},
  {Thongthai}, and {Swisdak}}}]{drake05}
\bibinfo{author}{\bibfnamefont{J.~F.} \bibnamefont{{Drake}}},
  \bibinfo{author}{\bibfnamefont{M.~A.} \bibnamefont{{Shay}}},
  \bibinfo{author}{\bibfnamefont{W.}~\bibnamefont{{Thongthai}}},
  \bibnamefont{and}
  \bibinfo{author}{\bibfnamefont{M.}~\bibnamefont{{Swisdak}}},
  \bibinfo{journal}{Phys. Rev. Lett.} \textbf{\bibinfo{volume}{94}},
  \bibinfo{pages}{095001} (\bibinfo{year}{2005}).

\bibitem[{\citenamefont{{Drake} et~al.}(2006)\citenamefont{{Drake}, {Swisdak},
  {Che}, and {Shay}}}]{drake06}
\bibinfo{author}{\bibfnamefont{J.~F.} \bibnamefont{{Drake}}},
  \bibinfo{author}{\bibfnamefont{M.}~\bibnamefont{{Swisdak}}},
  \bibinfo{author}{\bibfnamefont{H.}~\bibnamefont{{Che}}}, \bibnamefont{and}
  \bibinfo{author}{\bibfnamefont{M.~A.} \bibnamefont{{Shay}}},
  \bibinfo{journal}{Nature} \textbf{\bibinfo{volume}{443}},
  \bibinfo{pages}{553} (\bibinfo{year}{2006}).

\bibitem[{\citenamefont{{Brandt}}(1991)}]{brandt91}
\bibinfo{author}{\bibfnamefont{E.~H.} \bibnamefont{{Brandt}}},
  \bibinfo{journal}{Int. J. of Mod. Phys. B} \textbf{\bibinfo{volume}{5}},
  \bibinfo{pages}{751} (\bibinfo{year}{1991}).

\bibitem[{\citenamefont{{Bou-Diab} et~al.}(2001)\citenamefont{{Bou-Diab},
  {Dodgson}, and {Blatter}}}]{bou-diab01}
\bibinfo{author}{\bibfnamefont{M.}~\bibnamefont{{Bou-Diab}}},
  \bibinfo{author}{\bibfnamefont{M.~J.} \bibnamefont{{Dodgson}}},
  \bibnamefont{and}
  \bibinfo{author}{\bibfnamefont{G.}~\bibnamefont{{Blatter}}},
  \bibinfo{journal}{Phys. Rev. Lett.} \textbf{\bibinfo{volume}{86}},
  \bibinfo{pages}{5132} (\bibinfo{year}{2001}).

\bibitem[{\citenamefont{{Hindmarsh} and {Kibble}}(1995)}]{Hindmarsh95}
\bibinfo{author}{\bibfnamefont{M.~B.} \bibnamefont{{Hindmarsh}}}
  \bibnamefont{and} \bibinfo{author}{\bibfnamefont{T.~W.~B.}
  \bibnamefont{{Kibble}}}, \bibinfo{journal}{Rep. Prog. Phys.}
  \textbf{\bibinfo{volume}{58}}, \bibinfo{pages}{477} (\bibinfo{year}{1995}).

\bibitem[{\citenamefont{Fohl and Turner}(1975)}]{fohl75}
\bibinfo{author}{\bibfnamefont{T.}~\bibnamefont{Fohl}} \bibnamefont{and}
  \bibinfo{author}{\bibfnamefont{J.~S.} \bibnamefont{Turner}},
  \bibinfo{journal}{Phys. Fluids} \textbf{\bibinfo{volume}{18}},
  \bibinfo{pages}{433} (\bibinfo{year}{1975}).

\bibitem[{\citenamefont{{Ashurst} and {Meiron}}(1987)}]{ashurst87}
\bibinfo{author}{\bibfnamefont{W.~T.} \bibnamefont{{Ashurst}}}
  \bibnamefont{and} \bibinfo{author}{\bibfnamefont{D.~I.}
  \bibnamefont{{Meiron}}}, \bibinfo{journal}{Phys. Rev. Lett.}
  \textbf{\bibinfo{volume}{58}}, \bibinfo{pages}{1632} (\bibinfo{year}{1987}).

\bibitem[{\citenamefont{{Kerr} and {Hussain}}(1989)}]{kerr89}
\bibinfo{author}{\bibfnamefont{R.~M.} \bibnamefont{{Kerr}}} \bibnamefont{and}
  \bibinfo{author}{\bibfnamefont{F.}~\bibnamefont{{Hussain}}},
  \bibinfo{journal}{Physica D} \textbf{\bibinfo{volume}{37}},
  \bibinfo{pages}{474} (\bibinfo{year}{1989}).

\bibitem[{\citenamefont{{Siggia}}(1985)}]{siggia85}
\bibinfo{author}{\bibfnamefont{E.~D.} \bibnamefont{{Siggia}}},
  \bibinfo{journal}{Phys. Fluids} \textbf{\bibinfo{volume}{28}},
  \bibinfo{pages}{794} (\bibinfo{year}{1985}).

\bibitem[{\citenamefont{{Caradoc-Davies}
  et~al.}(2000)\citenamefont{{Caradoc-Davies}, {Ballagh}, and
  {Blakie}}}]{caradoc00}
\bibinfo{author}{\bibfnamefont{B.~M.} \bibnamefont{{Caradoc-Davies}}},
  \bibinfo{author}{\bibfnamefont{R.~J.} \bibnamefont{{Ballagh}}},
  \bibnamefont{and} \bibinfo{author}{\bibfnamefont{P.~B.}
  \bibnamefont{{Blakie}}}, \bibinfo{journal}{Phys. Rev. A}
  \textbf{\bibinfo{volume}{62}}, \bibinfo{pages}{011602 R}
  (\bibinfo{year}{2000}).

\bibitem[{\citenamefont{Feynman}(1955)}]{feynman55}
\bibinfo{author}{\bibfnamefont{R.~P.} \bibnamefont{Feynman}}, in
  \emph{\bibinfo{booktitle}{Progress in Low Temperature Physics}}, edited by
  \bibinfo{editor}{\bibfnamefont{C.~J.} \bibnamefont{Gorter}}
  (\bibinfo{publisher}{North-Holland}, \bibinfo{address}{Amsterdam},
  \bibinfo{year}{1955}), vol.~\bibinfo{volume}{1}, pp. \bibinfo{pages}{17--53}.

\bibitem[{\citenamefont{{Schwarz}}(1985)}]{schwarz85}
\bibinfo{author}{\bibfnamefont{K.~W.} \bibnamefont{{Schwarz}}},
  \bibinfo{journal}{Phys. Rev. B} \textbf{\bibinfo{volume}{31}},
  \bibinfo{pages}{5782} (\bibinfo{year}{1985}).

\bibitem[{\citenamefont{{Schwarz}}(1988)}]{schwarz88}
\bibinfo{author}{\bibfnamefont{K.~W.} \bibnamefont{{Schwarz}}},
  \bibinfo{journal}{Phys. Rev. B} \textbf{\bibinfo{volume}{38}},
  \bibinfo{pages}{2398} (\bibinfo{year}{1988}).

\bibitem[{\citenamefont{{Koplik} and {Levine}}(1993)}]{koplik93}
\bibinfo{author}{\bibfnamefont{J.}~\bibnamefont{{Koplik}}} \bibnamefont{and}
  \bibinfo{author}{\bibfnamefont{H.}~\bibnamefont{{Levine}}},
  \bibinfo{journal}{Phys. Rev. Lett.} \textbf{\bibinfo{volume}{71}},
  \bibinfo{pages}{1375} (\bibinfo{year}{1993}).

\bibitem[{\citenamefont{{de Waele} and {Aarts}}(1994)}]{deWaele94}
\bibinfo{author}{\bibfnamefont{A.~T.~A.~M.} \bibnamefont{{de Waele}}}
  \bibnamefont{and} \bibinfo{author}{\bibfnamefont{R.~G.~K.~M.}
  \bibnamefont{{Aarts}}}, \bibinfo{journal}{Phys. Rev. Lett.}
  \textbf{\bibinfo{volume}{72}}, \bibinfo{pages}{482} (\bibinfo{year}{1994}).

\bibitem[{\citenamefont{Gabbay et~al.}(1998)\citenamefont{Gabbay, Ott, and
  Guzdar}}]{gabbay98}
\bibinfo{author}{\bibfnamefont{M.}~\bibnamefont{Gabbay}},
  \bibinfo{author}{\bibfnamefont{E.}~\bibnamefont{Ott}}, \bibnamefont{and}
  \bibinfo{author}{\bibfnamefont{P.~N.} \bibnamefont{Guzdar}},
  \bibinfo{journal}{Phys. Rev. E} \textbf{\bibinfo{volume}{58}},
  \bibinfo{pages}{2576} (\bibinfo{year}{1998}).

\bibitem[{\citenamefont{{Lipniacki}}(2000)}]{lipniacki00}
\bibinfo{author}{\bibfnamefont{T.}~\bibnamefont{{Lipniacki}}},
  \bibinfo{journal}{Eur. J. Mech. B-Fluids} \textbf{\bibinfo{volume}{19}},
  \bibinfo{pages}{361} (\bibinfo{year}{2000}).

\bibitem[{\citenamefont{{Kivotides} et~al.}(2001)\citenamefont{{Kivotides},
  {Barenghi}, and {Samuels}}}]{kivotides01}
\bibinfo{author}{\bibfnamefont{D.}~\bibnamefont{{Kivotides}}},
  \bibinfo{author}{\bibfnamefont{C.~F.} \bibnamefont{{Barenghi}}},
  \bibnamefont{and} \bibinfo{author}{\bibfnamefont{D.~C.}
  \bibnamefont{{Samuels}}}, \bibinfo{journal}{Europhys. Lett.}
  \textbf{\bibinfo{volume}{54}}, \bibinfo{pages}{774} (\bibinfo{year}{2001}).

\bibitem[{\citenamefont{{Leadbeater} et~al.}(2001)\citenamefont{{Leadbeater},
  {Winiecki}, {Samuels}, {Barenghi}, and {Adams}}}]{leadbeater01}
\bibinfo{author}{\bibfnamefont{M.}~\bibnamefont{{Leadbeater}}},
  \bibinfo{author}{\bibfnamefont{T.}~\bibnamefont{{Winiecki}}},
  \bibinfo{author}{\bibfnamefont{D.~C.} \bibnamefont{{Samuels}}},
  \bibinfo{author}{\bibfnamefont{C.~F.} \bibnamefont{{Barenghi}}},
  \bibnamefont{and} \bibinfo{author}{\bibfnamefont{C.~S.}
  \bibnamefont{{Adams}}}, \bibinfo{journal}{Phys. Rev. Lett.}
  \textbf{\bibinfo{volume}{86}}, \bibinfo{pages}{1410} (\bibinfo{year}{2001}).

\bibitem[{\citenamefont{Vinen}(2001)}]{vinen01}
\bibinfo{author}{\bibfnamefont{W.~F.} \bibnamefont{Vinen}},
  \bibinfo{journal}{Phys. Rev. B} \textbf{\bibinfo{volume}{64}},
  \bibinfo{pages}{134520} (\bibinfo{year}{2001}).

\bibitem[{\citenamefont{{Ogawa} et~al.}(2002)\citenamefont{{Ogawa}, {Tsubota},
  and {Hattori}}}]{ogawa02}
\bibinfo{author}{\bibfnamefont{S.-i.} \bibnamefont{{Ogawa}}},
  \bibinfo{author}{\bibfnamefont{M.}~\bibnamefont{{Tsubota}}},
  \bibnamefont{and}
  \bibinfo{author}{\bibfnamefont{Y.}~\bibnamefont{{Hattori}}},
  \bibinfo{journal}{J. Phys. Soc. Jpn.} \textbf{\bibinfo{volume}{71}},
  \bibinfo{pages}{813} (\bibinfo{year}{2002}).

\bibitem[{\citenamefont{{Nazarenko} and {West}}(2003)}]{nazarenko03}
\bibinfo{author}{\bibfnamefont{S.}~\bibnamefont{{Nazarenko}}} \bibnamefont{and}
  \bibinfo{author}{\bibfnamefont{R.~J.} \bibnamefont{{West}}},
  \bibinfo{journal}{J. Low Temp. Phys.} \textbf{\bibinfo{volume}{132}},
  \bibinfo{pages}{1} (\bibinfo{year}{2003}).

\bibitem[{\citenamefont{{Vinen}}(2005)}]{vinen05}
\bibinfo{author}{\bibfnamefont{W.~F.} \bibnamefont{{Vinen}}},
  \bibinfo{journal}{J. Phys. Cond. Matter} \textbf{\bibinfo{volume}{17}},
  \bibinfo{pages}{3231} (\bibinfo{year}{2005}).

\bibitem[{\citenamefont{{Kuz'min}}(2006)}]{kuzmin06}
\bibinfo{author}{\bibfnamefont{P.~A.} \bibnamefont{{Kuz'min}}},
  \bibinfo{journal}{JETP Lett.} \textbf{\bibinfo{volume}{84}},
  \bibinfo{pages}{204} (\bibinfo{year}{2006}).

\bibitem[{\citenamefont{{Bewley} et~al.}(2008)\citenamefont{{Bewley},
  {Paoletti}, {Sreenivasan}, and {Lathrop}}}]{bewley08}
\bibinfo{author}{\bibfnamefont{G.~P.} \bibnamefont{{Bewley}}},
  \bibinfo{author}{\bibfnamefont{M.~S.} \bibnamefont{{Paoletti}}},
  \bibinfo{author}{\bibfnamefont{K.~R.} \bibnamefont{{Sreenivasan}}},
  \bibnamefont{and} \bibinfo{author}{\bibfnamefont{D.~P.}
  \bibnamefont{{Lathrop}}}, \bibinfo{journal}{Proc. Natl. Acad. Sci. U.S.A.}
  \textbf{\bibinfo{volume}{105}}, \bibinfo{pages}{13707}
  (\bibinfo{year}{2008}).

\bibitem[{\citenamefont{{Paoletti}
  et~al.}(2008{\natexlab{a}})\citenamefont{{Paoletti}, {Fisher}, {Sreenivasan},
  and {Lathrop}}}]{paoletti08a}
\bibinfo{author}{\bibfnamefont{M.~S.} \bibnamefont{{Paoletti}}},
  \bibinfo{author}{\bibfnamefont{M.~E.} \bibnamefont{{Fisher}}},
  \bibinfo{author}{\bibfnamefont{K.~R.} \bibnamefont{{Sreenivasan}}},
  \bibnamefont{and} \bibinfo{author}{\bibfnamefont{D.~P.}
  \bibnamefont{{Lathrop}}}, \bibinfo{journal}{Phys. Rev. Lett.}
  \textbf{\bibinfo{volume}{101}}, \bibinfo{pages}{154501}
  (\bibinfo{year}{2008}{\natexlab{a}}).

\bibitem[{\citenamefont{{Vinen}}(1957{\natexlab{a}})}]{vinen57a}
\bibinfo{author}{\bibfnamefont{W.~F.} \bibnamefont{{Vinen}}},
  \bibinfo{journal}{Proc. Roy. Soc. A} \textbf{\bibinfo{volume}{240}},
  \bibinfo{pages}{114} (\bibinfo{year}{1957}{\natexlab{a}}).

\bibitem[{\citenamefont{{Vinen}}(1957{\natexlab{b}})}]{vinen57b}
\bibinfo{author}{\bibfnamefont{W.~F.} \bibnamefont{{Vinen}}},
  \bibinfo{journal}{Proc. Roy. Soc. A} \textbf{\bibinfo{volume}{240}},
  \bibinfo{pages}{128} (\bibinfo{year}{1957}{\natexlab{b}}).

\bibitem[{\citenamefont{{Vinen}}(1957{\natexlab{c}})}]{vinen57c}
\bibinfo{author}{\bibfnamefont{W.~F.} \bibnamefont{{Vinen}}},
  \bibinfo{journal}{Proc. Roy. Soc. A} \textbf{\bibinfo{volume}{242}},
  \bibinfo{pages}{493} (\bibinfo{year}{1957}{\natexlab{c}}).

\bibitem[{\citenamefont{{Walmsley} et~al.}(2007)\citenamefont{{Walmsley},
  {Golov}, {Hall}, {Levchenko}, and {Vinen}}}]{walmsley07}
\bibinfo{author}{\bibfnamefont{P.~M.} \bibnamefont{{Walmsley}}},
  \bibinfo{author}{\bibfnamefont{A.~I.} \bibnamefont{{Golov}}},
  \bibinfo{author}{\bibfnamefont{H.~E.} \bibnamefont{{Hall}}},
  \bibinfo{author}{\bibfnamefont{A.~A.} \bibnamefont{{Levchenko}}},
  \bibnamefont{and} \bibinfo{author}{\bibfnamefont{W.~F.}
  \bibnamefont{{Vinen}}}, \bibinfo{journal}{Phys. Rev. Lett.}
  \textbf{\bibinfo{volume}{99}}, \bibinfo{eid}{265302} (\bibinfo{year}{2007}).

\bibitem[{\citenamefont{{Poole} et~al.}(2005)\citenamefont{{Poole}, {Barenghi},
  {Sergeev}, and {Vinen}}}]{poole05}
\bibinfo{author}{\bibfnamefont{D.~R.} \bibnamefont{{Poole}}},
  \bibinfo{author}{\bibfnamefont{C.~F.} \bibnamefont{{Barenghi}}},
  \bibinfo{author}{\bibfnamefont{Y.~A.} \bibnamefont{{Sergeev}}},
  \bibnamefont{and} \bibinfo{author}{\bibfnamefont{W.~F.}
  \bibnamefont{{Vinen}}}, \bibinfo{journal}{Phys. Rev. B}
  \textbf{\bibinfo{volume}{71}}, \bibinfo{eid}{064514} (\bibinfo{year}{2005}).

\bibitem[{\citenamefont{{Bewley} et~al.}(2006)\citenamefont{{Bewley},
  {Lathrop}, and {Sreenivasan}}}]{bewley06}
\bibinfo{author}{\bibfnamefont{G.~P.} \bibnamefont{{Bewley}}},
  \bibinfo{author}{\bibfnamefont{D.~P.} \bibnamefont{{Lathrop}}},
  \bibnamefont{and} \bibinfo{author}{\bibfnamefont{K.~R.}
  \bibnamefont{{Sreenivasan}}}, \bibinfo{journal}{Nature}
  \textbf{\bibinfo{volume}{441}}, \bibinfo{pages}{588} (\bibinfo{year}{2006}).

\bibitem[{\citenamefont{{Schwarz}}(1978)}]{schwarz78}
\bibinfo{author}{\bibfnamefont{K.~W.} \bibnamefont{{Schwarz}}},
  \bibinfo{journal}{Phys. Rev. B} \textbf{\bibinfo{volume}{18}},
  \bibinfo{pages}{245} (\bibinfo{year}{1978}).

\bibitem[{\citenamefont{{Paoletti}
  et~al.}(2008{\natexlab{b}})\citenamefont{{Paoletti}, {Fiorito},
  {Sreenivasan}, and {Lathrop}}}]{paoletti08b}
\bibinfo{author}{\bibfnamefont{M.~S.} \bibnamefont{{Paoletti}}},
  \bibinfo{author}{\bibfnamefont{R.~B.} \bibnamefont{{Fiorito}}},
  \bibinfo{author}{\bibfnamefont{K.~R.} \bibnamefont{{Sreenivasan}}},
  \bibnamefont{and} \bibinfo{author}{\bibfnamefont{D.~P.}
  \bibnamefont{{Lathrop}}}, \bibinfo{journal}{J. Phys. Soc. Jpn.}
  \textbf{\bibinfo{volume}{77}}, \bibinfo{pages}{in press}
  (\bibinfo{year}{2008}{\natexlab{b}}).

\bibitem[{\citenamefont{{Skrbek} et~al.}(2003)\citenamefont{{Skrbek},
  {Gordeev}, and {Soukup}}}]{skrbek03}
\bibinfo{author}{\bibfnamefont{L.}~\bibnamefont{{Skrbek}}},
  \bibinfo{author}{\bibfnamefont{A.~V.} \bibnamefont{{Gordeev}}},
  \bibnamefont{and} \bibinfo{author}{\bibfnamefont{F.}~\bibnamefont{{Soukup}}},
  \bibinfo{journal}{Phys. Rev. E} \textbf{\bibinfo{volume}{67}},
  \bibinfo{pages}{047302} (\bibinfo{year}{2003}).

\bibitem[{wee()}]{weeks}
\bibinfo{note}{We thank Eric Weeks and John Crocker for the particle-tracking
  algorithm.}

\bibitem[{\citenamefont{Adrian and Yao}(1984)}]{adrian84}
\bibinfo{author}{\bibfnamefont{R.~J.} \bibnamefont{Adrian}} \bibnamefont{and}
  \bibinfo{author}{\bibfnamefont{C.~S.} \bibnamefont{Yao}}, in
  \emph{\bibinfo{booktitle}{Proceedings, Eigth Biennial Symposium on
  Turbulence}}, edited by
  \bibinfo{editor}{\bibfnamefont{G.}~\bibnamefont{Patterson}} \bibnamefont{and}
  \bibinfo{editor}{\bibfnamefont{J.~L.} \bibnamefont{Zakin}}
  (\bibinfo{publisher}{U. Missouri}, \bibinfo{address}{Rolla},
  \bibinfo{year}{1984}), pp. \bibinfo{pages}{170--186}.

\bibitem[{\citenamefont{{Sreenivasan} and {Meneveau}}(1988)}]{sreenivasan88}
\bibinfo{author}{\bibfnamefont{K.~R.} \bibnamefont{{Sreenivasan}}}
  \bibnamefont{and}
  \bibinfo{author}{\bibfnamefont{C.}~\bibnamefont{{Meneveau}}},
  \bibinfo{journal}{Phys. Rev. A} \textbf{\bibinfo{volume}{38}},
  \bibinfo{pages}{6287} (\bibinfo{year}{1988}).

\bibitem[{\citenamefont{Wegner}(1972)}]{wegner72}
\bibinfo{author}{\bibfnamefont{F.~J.} \bibnamefont{Wegner}},
  \bibinfo{journal}{Phys. Rev. B} \textbf{\bibinfo{volume}{5}},
  \bibinfo{pages}{4529} (\bibinfo{year}{1972}).

\bibitem[{\citenamefont{Chen et~al.}(1982)\citenamefont{Chen, Fisher, and
  Nickel}}]{chen82}
\bibinfo{author}{\bibfnamefont{J.-H.} \bibnamefont{Chen}},
  \bibinfo{author}{\bibfnamefont{M.~E.} \bibnamefont{Fisher}},
  \bibnamefont{and} \bibinfo{author}{\bibfnamefont{B.~G.}
  \bibnamefont{Nickel}}, \bibinfo{journal}{Phys. Rev. Lett.}
  \textbf{\bibinfo{volume}{48}}, \bibinfo{pages}{630} (\bibinfo{year}{1982}).

\bibitem[{\citenamefont{{Guida} and {Zinn-Justin}}(1997)}]{guida97}
\bibinfo{author}{\bibfnamefont{R.}~\bibnamefont{{Guida}}} \bibnamefont{and}
  \bibinfo{author}{\bibfnamefont{J.}~\bibnamefont{{Zinn-Justin}}},
  \bibinfo{journal}{Nucl. Phys. B} \textbf{\bibinfo{volume}{489}},
  \bibinfo{pages}{626} (\bibinfo{year}{1997}).

\bibitem[{\citenamefont{Wang et~al.}(1987)\citenamefont{Wang, Swanson, and
  Donnelly}}]{wang87}
\bibinfo{author}{\bibfnamefont{R.~T.} \bibnamefont{Wang}},
  \bibinfo{author}{\bibfnamefont{C.~E.} \bibnamefont{Swanson}},
  \bibnamefont{and} \bibinfo{author}{\bibfnamefont{R.~J.}
  \bibnamefont{Donnelly}}, \bibinfo{journal}{Phys. Rev. B}
  \textbf{\bibinfo{volume}{36}}, \bibinfo{pages}{5240} (\bibinfo{year}{1987}).

\bibitem[{\citenamefont{{Gordeev} et~al.}(2005)\citenamefont{{Gordeev},
  {Chagovets}, {Soukup}, and {Skrbek}}}]{gordeev05}
\bibinfo{author}{\bibfnamefont{A.~V.} \bibnamefont{{Gordeev}}},
  \bibinfo{author}{\bibfnamefont{T.~V.} \bibnamefont{{Chagovets}}},
  \bibinfo{author}{\bibfnamefont{F.}~\bibnamefont{{Soukup}}}, \bibnamefont{and}
  \bibinfo{author}{\bibfnamefont{L.}~\bibnamefont{{Skrbek}}},
  \bibinfo{journal}{J. Low Temp. Phys.} \textbf{\bibinfo{volume}{138}},
  \bibinfo{pages}{549} (\bibinfo{year}{2005}).

\bibitem[{\citenamefont{{Barenghi} et~al.}(2006)\citenamefont{{Barenghi},
  {Gordeev}, and {Skrbek}}}]{barenghi06}
\bibinfo{author}{\bibfnamefont{C.~F.} \bibnamefont{{Barenghi}}},
  \bibinfo{author}{\bibfnamefont{A.~V.} \bibnamefont{{Gordeev}}},
  \bibnamefont{and} \bibinfo{author}{\bibfnamefont{L.}~\bibnamefont{{Skrbek}}},
  \bibinfo{journal}{Phys. Rev. E} \textbf{\bibinfo{volume}{74}},
  \bibinfo{pages}{026309} (\bibinfo{year}{2006}).

\bibitem[{\citenamefont{{Barenghi} and {Skrbek}}(2007)}]{barenghi07}
\bibinfo{author}{\bibfnamefont{C.~F.} \bibnamefont{{Barenghi}}}
  \bibnamefont{and} \bibinfo{author}{\bibfnamefont{L.}~\bibnamefont{{Skrbek}}},
  \bibinfo{journal}{J. Low Temp. Phys.} \textbf{\bibinfo{volume}{146}},
  \bibinfo{pages}{5} (\bibinfo{year}{2007}).

\bibitem[{\citenamefont{Vainshtein and Cattaneo}(1992)}]{vainshtein92}
\bibinfo{author}{\bibfnamefont{S.}~\bibnamefont{Vainshtein}} \bibnamefont{and}
  \bibinfo{author}{\bibfnamefont{F.}~\bibnamefont{Cattaneo}},
  \bibinfo{journal}{Astrophys. J.} \textbf{\bibinfo{volume}{393}},
  \bibinfo{pages}{165} (\bibinfo{year}{1992}).

\bibitem[{\citenamefont{Balbus and Hawley}(1998)}]{balbus98}
\bibinfo{author}{\bibfnamefont{S.~A.} \bibnamefont{Balbus}} \bibnamefont{and}
  \bibinfo{author}{\bibfnamefont{J.~F.} \bibnamefont{Hawley}},
  \bibinfo{journal}{Rev. Mod. Phys.} \textbf{\bibinfo{volume}{70}},
  \bibinfo{pages}{1} (\bibinfo{year}{1998}).

\bibitem[{\citenamefont{{Smith} et~al.}(1993)\citenamefont{{Smith}, {Donnelly},
  {Goldenfeld}, and {Vinen}}}]{smith93}
\bibinfo{author}{\bibfnamefont{M.~R.} \bibnamefont{{Smith}}},
  \bibinfo{author}{\bibfnamefont{R.~J.} \bibnamefont{{Donnelly}}},
  \bibinfo{author}{\bibfnamefont{N.}~\bibnamefont{{Goldenfeld}}},
  \bibnamefont{and} \bibinfo{author}{\bibfnamefont{W.~F.}
  \bibnamefont{{Vinen}}}, \bibinfo{journal}{Phys. Rev. Lett.}
  \textbf{\bibinfo{volume}{71}}, \bibinfo{pages}{2583} (\bibinfo{year}{1993}).

\bibitem[{\citenamefont{{Barenghi} et~al.}(1997)\citenamefont{{Barenghi},
  {Samuels}, {Bauer}, and {Donnelly}}}]{barenghi97}
\bibinfo{author}{\bibfnamefont{C.~F.} \bibnamefont{{Barenghi}}},
  \bibinfo{author}{\bibfnamefont{D.~C.} \bibnamefont{{Samuels}}},
  \bibinfo{author}{\bibfnamefont{G.~H.} \bibnamefont{{Bauer}}},
  \bibnamefont{and} \bibinfo{author}{\bibfnamefont{R.~J.}
  \bibnamefont{{Donnelly}}}, \bibinfo{journal}{Phys. Fluids}
  \textbf{\bibinfo{volume}{9}}, \bibinfo{pages}{2631} (\bibinfo{year}{1997}).

\bibitem[{\citenamefont{{Nore} et~al.}(1997)\citenamefont{{Nore}, {Abid}, and
  {Brachet}}}]{nore97}
\bibinfo{author}{\bibfnamefont{C.}~\bibnamefont{{Nore}}},
  \bibinfo{author}{\bibfnamefont{M.}~\bibnamefont{{Abid}}}, \bibnamefont{and}
  \bibinfo{author}{\bibfnamefont{M.~E.} \bibnamefont{{Brachet}}},
  \bibinfo{journal}{Phys. Rev. Lett.} \textbf{\bibinfo{volume}{78}},
  \bibinfo{pages}{3896} (\bibinfo{year}{1997}).

\bibitem[{\citenamefont{{Maurer} and {Tabeling}}(1998)}]{maurer98}
\bibinfo{author}{\bibfnamefont{J.}~\bibnamefont{{Maurer}}} \bibnamefont{and}
  \bibinfo{author}{\bibfnamefont{P.}~\bibnamefont{{Tabeling}}},
  \bibinfo{journal}{Europhys. Lett.} \textbf{\bibinfo{volume}{43}},
  \bibinfo{pages}{29} (\bibinfo{year}{1998}).

\bibitem[{\citenamefont{{Barenghi}}(1999)}]{barenghi99}
\bibinfo{author}{\bibfnamefont{C.~F.} \bibnamefont{{Barenghi}}},
  \bibinfo{journal}{J. Phys. Cond. Matter} \textbf{\bibinfo{volume}{11}},
  \bibinfo{pages}{7751} (\bibinfo{year}{1999}).

\bibitem[{\citenamefont{{Stalp} et~al.}(1999)\citenamefont{{Stalp}, {Skrbek},
  and {Donnelly}}}]{stalp99}
\bibinfo{author}{\bibfnamefont{S.~R.} \bibnamefont{{Stalp}}},
  \bibinfo{author}{\bibfnamefont{L.}~\bibnamefont{{Skrbek}}}, \bibnamefont{and}
  \bibinfo{author}{\bibfnamefont{R.~J.} \bibnamefont{{Donnelly}}},
  \bibinfo{journal}{Phys. Rev. Lett.} \textbf{\bibinfo{volume}{82}},
  \bibinfo{pages}{4831} (\bibinfo{year}{1999}).

\bibitem[{\citenamefont{{Vinen}}(2000)}]{vinen00}
\bibinfo{author}{\bibfnamefont{W.~F.} \bibnamefont{{Vinen}}},
  \bibinfo{journal}{Phys. Rev. B} \textbf{\bibinfo{volume}{61}},
  \bibinfo{pages}{1410} (\bibinfo{year}{2000}).

\bibitem[{\citenamefont{{Skrbek} et~al.}(2000)\citenamefont{{Skrbek},
  {Niemela}, and {Donnelly}}}]{skrbek00}
\bibinfo{author}{\bibfnamefont{L.}~\bibnamefont{{Skrbek}}},
  \bibinfo{author}{\bibfnamefont{J.~J.} \bibnamefont{{Niemela}}},
  \bibnamefont{and} \bibinfo{author}{\bibfnamefont{R.~J.}
  \bibnamefont{{Donnelly}}}, \bibinfo{journal}{Phys. Rev. Lett.}
  \textbf{\bibinfo{volume}{85}}, \bibinfo{pages}{2973} (\bibinfo{year}{2000}).

\bibitem[{\citenamefont{Vinen and J.}(2002)}]{vinen02}
\bibinfo{author}{\bibfnamefont{W.~F.} \bibnamefont{Vinen}} \bibnamefont{and}
  \bibinfo{author}{\bibfnamefont{N.~J.} \bibnamefont{J.}}, \bibinfo{journal}{J.
  Low Temp. Phys.} \textbf{\bibinfo{volume}{128}}, \bibinfo{pages}{167}
  (\bibinfo{year}{2002}).

\bibitem[{\citenamefont{Barenghi et~al.}(2002)\citenamefont{Barenghi, Hulton,
  and Samuels}}]{barenghi02a}
\bibinfo{author}{\bibfnamefont{C.~F.} \bibnamefont{Barenghi}},
  \bibinfo{author}{\bibfnamefont{S.}~\bibnamefont{Hulton}}, \bibnamefont{and}
  \bibinfo{author}{\bibfnamefont{D.~C.} \bibnamefont{Samuels}},
  \bibinfo{journal}{Phys. Rev. Lett.} \textbf{\bibinfo{volume}{89}},
  \bibinfo{pages}{275301} (\bibinfo{year}{2002}).

\bibitem[{\citenamefont{{Barenghi} et~al.}(2002)\citenamefont{{Barenghi},
  {Samuels}, and {Kivotides}}}]{barenghi02b}
\bibinfo{author}{\bibfnamefont{C.~F.} \bibnamefont{{Barenghi}}},
  \bibinfo{author}{\bibfnamefont{D.~C.} \bibnamefont{{Samuels}}},
  \bibnamefont{and}
  \bibinfo{author}{\bibfnamefont{D.}~\bibnamefont{{Kivotides}}},
  \bibinfo{journal}{J. Low Temp. Phys.} \textbf{\bibinfo{volume}{126}},
  \bibinfo{pages}{271} (\bibinfo{year}{2002}).

\bibitem[{\citenamefont{{Kobayashi} and {Tsubota}}(2005)}]{kobayashi05}
\bibinfo{author}{\bibfnamefont{M.}~\bibnamefont{{Kobayashi}}} \bibnamefont{and}
  \bibinfo{author}{\bibfnamefont{M.}~\bibnamefont{{Tsubota}}},
  \bibinfo{journal}{Phys. Rev. Lett.} \textbf{\bibinfo{volume}{94}},
  \bibinfo{pages}{065302} (\bibinfo{year}{2005}).

\bibitem[{\citenamefont{{Kobayashi} and {Tsubota}}(2006)}]{kobayashi06}
\bibinfo{author}{\bibfnamefont{M.}~\bibnamefont{{Kobayashi}}} \bibnamefont{and}
  \bibinfo{author}{\bibfnamefont{M.}~\bibnamefont{{Tsubota}}},
  \bibinfo{journal}{J. Low Temp. Phys.} \textbf{\bibinfo{volume}{145}},
  \bibinfo{pages}{209} (\bibinfo{year}{2006}).

\bibitem[{\citenamefont{{Kobayashi} and {Tsubota}}(2007)}]{kobayashi07}
\bibinfo{author}{\bibfnamefont{M.}~\bibnamefont{{Kobayashi}}} \bibnamefont{and}
  \bibinfo{author}{\bibfnamefont{M.}~\bibnamefont{{Tsubota}}},
  \bibinfo{journal}{Phys. Rev. A} \textbf{\bibinfo{volume}{76}},
  \bibinfo{pages}{045603} (\bibinfo{year}{2007}).

\bibitem[{\citenamefont{{L'Vov} et~al.}(2007)\citenamefont{{L'Vov},
  {Nazarenko}, and {Rudenko}}}]{l'vov07}
\bibinfo{author}{\bibfnamefont{V.~S.} \bibnamefont{{L'Vov}}},
  \bibinfo{author}{\bibfnamefont{S.~V.} \bibnamefont{{Nazarenko}}},
  \bibnamefont{and}
  \bibinfo{author}{\bibfnamefont{O.}~\bibnamefont{{Rudenko}}},
  \bibinfo{journal}{Phys. Rev. B} \textbf{\bibinfo{volume}{76}},
  \bibinfo{pages}{024520} (\bibinfo{year}{2007}).

\bibitem[{\citenamefont{Chagovets et~al.}(2007)\citenamefont{Chagovets,
  Gordeev, and Skrbek}}]{chagovets07}
\bibinfo{author}{\bibfnamefont{T.~V.} \bibnamefont{Chagovets}},
  \bibinfo{author}{\bibfnamefont{A.~V.} \bibnamefont{Gordeev}},
  \bibnamefont{and} \bibinfo{author}{\bibfnamefont{L.}~\bibnamefont{Skrbek}},
  \bibinfo{journal}{Phys. Rev. E} \textbf{\bibinfo{volume}{76}},
  \bibinfo{eid}{027301} (\bibinfo{year}{2007}).

\bibitem[{\citenamefont{Morris et~al.}(2008)\citenamefont{Morris, Koplik, and
  Rouson}}]{morris08}
\bibinfo{author}{\bibfnamefont{K.}~\bibnamefont{Morris}},
  \bibinfo{author}{\bibfnamefont{J.}~\bibnamefont{Koplik}}, \bibnamefont{and}
  \bibinfo{author}{\bibfnamefont{D.~W.~I.} \bibnamefont{Rouson}},
  \bibinfo{journal}{Phys. Rev. Lett.} \textbf{\bibinfo{volume}{101}},
  \bibinfo{eid}{015301} (\bibinfo{year}{2008}).

\bibitem[{\citenamefont{{Procaccia} and {Sreenivasan}}(2008)}]{procaccia08}
\bibinfo{author}{\bibfnamefont{I.}~\bibnamefont{{Procaccia}}} \bibnamefont{and}
  \bibinfo{author}{\bibfnamefont{K.~R.} \bibnamefont{{Sreenivasan}}},
  \bibinfo{journal}{Physica D} \textbf{\bibinfo{volume}{237}},
  \bibinfo{pages}{2167} (\bibinfo{year}{2008}).

\bibitem[{\citenamefont{Frisch}(1995)}]{frisch95}
\bibinfo{author}{\bibfnamefont{U.}~\bibnamefont{Frisch}},
  \emph{\bibinfo{title}{Turbulence: The Legacy of A. N. Kolmogorov}}
  (\bibinfo{publisher}{Cambridge Univ. Press}, \bibinfo{address}{Cambridge,
  UK}, \bibinfo{year}{1995}).

\bibitem[{\citenamefont{{Noullez} et~al.}(1997)\citenamefont{{Noullez},
  {Wallace}, {Lempert}, {Miles}, and {Frisch}}}]{noullez97}
\bibinfo{author}{\bibfnamefont{A.}~\bibnamefont{{Noullez}}},
  \bibinfo{author}{\bibfnamefont{G.}~\bibnamefont{{Wallace}}},
  \bibinfo{author}{\bibfnamefont{W.}~\bibnamefont{{Lempert}}},
  \bibinfo{author}{\bibfnamefont{R.~B.} \bibnamefont{{Miles}}},
  \bibnamefont{and} \bibinfo{author}{\bibfnamefont{U.}~\bibnamefont{{Frisch}}},
  \bibinfo{journal}{J. Fluid Mech.} \textbf{\bibinfo{volume}{339}},
  \bibinfo{pages}{287} (\bibinfo{year}{1997}).

\bibitem[{\citenamefont{{Vincent} and {Meneguzzi}}(1991)}]{vincent91}
\bibinfo{author}{\bibfnamefont{A.}~\bibnamefont{{Vincent}}} \bibnamefont{and}
  \bibinfo{author}{\bibfnamefont{M.}~\bibnamefont{{Meneguzzi}}},
  \bibinfo{journal}{J. Fluid Mech.} \textbf{\bibinfo{volume}{225}},
  \bibinfo{pages}{1} (\bibinfo{year}{1991}).

\bibitem[{\citenamefont{{Gotoh} et~al.}(2002)\citenamefont{{Gotoh}, {Fukayama},
  and {Nakano}}}]{gotoh02}
\bibinfo{author}{\bibfnamefont{T.}~\bibnamefont{{Gotoh}}},
  \bibinfo{author}{\bibfnamefont{D.}~\bibnamefont{{Fukayama}}},
  \bibnamefont{and} \bibinfo{author}{\bibfnamefont{T.}~\bibnamefont{{Nakano}}},
  \bibinfo{journal}{Phys. Fluids} \textbf{\bibinfo{volume}{14}},
  \bibinfo{pages}{1065} (\bibinfo{year}{2002}).

\bibitem[{\citenamefont{{Kolmogorov}}(1941)}]{kolmogorov41a}
\bibinfo{author}{\bibfnamefont{A.}~\bibnamefont{{Kolmogorov}}},
  \bibinfo{journal}{Dokl. Akad. Nauk SSSR} \textbf{\bibinfo{volume}{30}},
  \bibinfo{pages}{301} (\bibinfo{year}{1941}), \bibinfo{note}{{[Proc. R. Soc.
  Lond. A \textbf{434,} 9 (1991)]}}.

\bibitem[{\citenamefont{{Kolmogorov}}(1962)}]{kolmogorov62}
\bibinfo{author}{\bibfnamefont{A.~N.} \bibnamefont{{Kolmogorov}}},
  \bibinfo{journal}{J. Fluid Mech.} \textbf{\bibinfo{volume}{13}},
  \bibinfo{pages}{82} (\bibinfo{year}{1962}).

\bibitem[{\citenamefont{{Benzi} et~al.}(1984)\citenamefont{{Benzi}, {Paladin},
  {Vulpiani}, and {Parisi}}}]{benzi84}
\bibinfo{author}{\bibfnamefont{R.}~\bibnamefont{{Benzi}}},
  \bibinfo{author}{\bibfnamefont{G.}~\bibnamefont{{Paladin}}},
  \bibinfo{author}{\bibfnamefont{A.}~\bibnamefont{{Vulpiani}}},
  \bibnamefont{and} \bibinfo{author}{\bibfnamefont{G.}~\bibnamefont{{Parisi}}},
  \bibinfo{journal}{J. Phys. A} \textbf{\bibinfo{volume}{17}},
  \bibinfo{pages}{3521} (\bibinfo{year}{1984}).

\bibitem[{\citenamefont{{Anselmet} et~al.}(1984)\citenamefont{{Anselmet},
  {Gagne}, {Hopfinger}, and {Antonia}}}]{anselmet84}
\bibinfo{author}{\bibfnamefont{F.}~\bibnamefont{{Anselmet}}},
  \bibinfo{author}{\bibfnamefont{Y.}~\bibnamefont{{Gagne}}},
  \bibinfo{author}{\bibfnamefont{E.~J.} \bibnamefont{{Hopfinger}}},
  \bibnamefont{and} \bibinfo{author}{\bibfnamefont{R.~A.}
  \bibnamefont{{Antonia}}}, \bibinfo{journal}{J. Fluid Mech.}
  \textbf{\bibinfo{volume}{140}}, \bibinfo{pages}{63} (\bibinfo{year}{1984}).

\bibitem[{\citenamefont{{Meneveau} and {Sreenivasan}}(1987)}]{meneveau87}
\bibinfo{author}{\bibfnamefont{C.}~\bibnamefont{{Meneveau}}} \bibnamefont{and}
  \bibinfo{author}{\bibfnamefont{K.~R.} \bibnamefont{{Sreenivasan}}},
  \bibinfo{journal}{Phys. Rev. Lett.} \textbf{\bibinfo{volume}{59}},
  \bibinfo{pages}{1424} (\bibinfo{year}{1987}).

\bibitem[{\citenamefont{{Andrews} et~al.}(1989)\citenamefont{{Andrews},
  {Phillips}, {Shivamoggi}, {Beck}, and {Joshi}}}]{andrews89}
\bibinfo{author}{\bibfnamefont{L.~C.} \bibnamefont{{Andrews}}},
  \bibinfo{author}{\bibfnamefont{R.~L.} \bibnamefont{{Phillips}}},
  \bibinfo{author}{\bibfnamefont{B.~K.} \bibnamefont{{Shivamoggi}}},
  \bibinfo{author}{\bibfnamefont{J.~K.} \bibnamefont{{Beck}}},
  \bibnamefont{and} \bibinfo{author}{\bibfnamefont{M.~L.}
  \bibnamefont{{Joshi}}}, \bibinfo{journal}{Phys. Fluids}
  \textbf{\bibinfo{volume}{1}}, \bibinfo{pages}{999} (\bibinfo{year}{1989}).

\bibitem[{\citenamefont{{Kida}}(1991)}]{kida91}
\bibinfo{author}{\bibfnamefont{S.}~\bibnamefont{{Kida}}}, \bibinfo{journal}{J.
  Phys. Soc. Jpn} \textbf{\bibinfo{volume}{60}}, \bibinfo{pages}{5}
  (\bibinfo{year}{1991}).

\bibitem[{\citenamefont{She and Orszag}(1991)}]{she91}
\bibinfo{author}{\bibfnamefont{Z.-S.} \bibnamefont{She}} \bibnamefont{and}
  \bibinfo{author}{\bibfnamefont{S.~A.} \bibnamefont{Orszag}},
  \bibinfo{journal}{Phys. Rev. Lett.} \textbf{\bibinfo{volume}{66}},
  \bibinfo{pages}{1701} (\bibinfo{year}{1991}).

\bibitem[{\citenamefont{{Stolovitzky} and {Sreenivasan}}(1993)}]{stolovitzky93}
\bibinfo{author}{\bibfnamefont{G.}~\bibnamefont{{Stolovitzky}}}
  \bibnamefont{and} \bibinfo{author}{\bibfnamefont{K.~R.}
  \bibnamefont{{Sreenivasan}}}, \bibinfo{journal}{Phys. Rev. E}
  \textbf{\bibinfo{volume}{48}}, \bibinfo{pages}{33} (\bibinfo{year}{1993}).

\bibitem[{\citenamefont{Benzi et~al.}(1993)\citenamefont{Benzi, Ciliberto,
  Tripiccione, Baudet, Massaioli, and Succi}}]{benzi93a}
\bibinfo{author}{\bibfnamefont{R.}~\bibnamefont{Benzi}},
  \bibinfo{author}{\bibfnamefont{S.}~\bibnamefont{Ciliberto}},
  \bibinfo{author}{\bibfnamefont{R.}~\bibnamefont{Tripiccione}},
  \bibinfo{author}{\bibfnamefont{C.}~\bibnamefont{Baudet}},
  \bibinfo{author}{\bibfnamefont{F.}~\bibnamefont{Massaioli}},
  \bibnamefont{and} \bibinfo{author}{\bibfnamefont{S.}~\bibnamefont{Succi}},
  \bibinfo{journal}{Phys. Rev. E} \textbf{\bibinfo{volume}{48}},
  \bibinfo{pages}{R29} (\bibinfo{year}{1993}).

\bibitem[{\citenamefont{{Benzi} et~al.}(1993)\citenamefont{{Benzi},
  {Ciliberto}, {Baudet}, {Ruiz Chavarria}, and {Tripiccione}}}]{benzi93b}
\bibinfo{author}{\bibfnamefont{R.}~\bibnamefont{{Benzi}}},
  \bibinfo{author}{\bibfnamefont{S.}~\bibnamefont{{Ciliberto}}},
  \bibinfo{author}{\bibfnamefont{C.}~\bibnamefont{{Baudet}}},
  \bibinfo{author}{\bibfnamefont{G.}~\bibnamefont{{Ruiz Chavarria}}},
  \bibnamefont{and}
  \bibinfo{author}{\bibfnamefont{R.}~\bibnamefont{{Tripiccione}}},
  \bibinfo{journal}{Europhys. Lett.} \textbf{\bibinfo{volume}{24}},
  \bibinfo{pages}{275} (\bibinfo{year}{1993}).

\bibitem[{\citenamefont{{She} and {Leveque}}(1994)}]{she94}
\bibinfo{author}{\bibfnamefont{Z.-S.} \bibnamefont{{She}}} \bibnamefont{and}
  \bibinfo{author}{\bibfnamefont{E.}~\bibnamefont{{Leveque}}},
  \bibinfo{journal}{Phys. Rev. Lett.} \textbf{\bibinfo{volume}{72}},
  \bibinfo{pages}{336} (\bibinfo{year}{1994}).

\bibitem[{\citenamefont{{Barenblatt} and {Goldenfeld}}(1995)}]{barenblatt95}
\bibinfo{author}{\bibfnamefont{G.~I.} \bibnamefont{{Barenblatt}}}
  \bibnamefont{and}
  \bibinfo{author}{\bibfnamefont{N.}~\bibnamefont{{Goldenfeld}}},
  \bibinfo{journal}{Phys. Fluids} \textbf{\bibinfo{volume}{7}},
  \bibinfo{pages}{3078} (\bibinfo{year}{1995}).

\bibitem[{\citenamefont{{Arneodo} et~al.}(1996)\citenamefont{{Arneodo},
  {Baudet}, {Belin}, {Benzi}, {Castaing}, {Chabaud}, {Chavarria}, {Ciliberto},
  {Camussi}, {Chill{\`a}} et~al.}}]{arneodo96}
\bibinfo{author}{\bibfnamefont{A.}~\bibnamefont{{Arneodo}}},
  \bibinfo{author}{\bibfnamefont{C.}~\bibnamefont{{Baudet}}},
  \bibinfo{author}{\bibfnamefont{F.}~\bibnamefont{{Belin}}},
  \bibinfo{author}{\bibfnamefont{R.}~\bibnamefont{{Benzi}}},
  \bibinfo{author}{\bibfnamefont{B.}~\bibnamefont{{Castaing}}},
  \bibinfo{author}{\bibfnamefont{B.}~\bibnamefont{{Chabaud}}},
  \bibinfo{author}{\bibfnamefont{R.}~\bibnamefont{{Chavarria}}},
  \bibinfo{author}{\bibfnamefont{S.}~\bibnamefont{{Ciliberto}}},
  \bibinfo{author}{\bibfnamefont{R.}~\bibnamefont{{Camussi}}},
  \bibinfo{author}{\bibfnamefont{F.}~\bibnamefont{{Chill{\`a}}}},
  \bibnamefont{et~al.}, \bibinfo{journal}{Europhys. Lett.}
  \textbf{\bibinfo{volume}{34}}, \bibinfo{pages}{411} (\bibinfo{year}{1996}).

\bibitem[{\citenamefont{{Boratav} and {Pelz}}(1997)}]{boratav97}
\bibinfo{author}{\bibfnamefont{O.~N.} \bibnamefont{{Boratav}}}
  \bibnamefont{and} \bibinfo{author}{\bibfnamefont{R.~B.}
  \bibnamefont{{Pelz}}}, \bibinfo{journal}{Physics of Fluids}
  \textbf{\bibinfo{volume}{9}}, \bibinfo{pages}{1400} (\bibinfo{year}{1997}).

\bibitem[{\citenamefont{{Sreenivasan} and {Antonia}}(1997)}]{sreenivasan97}
\bibinfo{author}{\bibfnamefont{K.~R.} \bibnamefont{{Sreenivasan}}}
  \bibnamefont{and} \bibinfo{author}{\bibfnamefont{R.~A.}
  \bibnamefont{{Antonia}}}, \bibinfo{journal}{Ann. Rev. Fluid Mech.}
  \textbf{\bibinfo{volume}{29}}, \bibinfo{pages}{435} (\bibinfo{year}{1997}).

\bibitem[{\citenamefont{Lewis and Swinney}(1999)}]{lewis99}
\bibinfo{author}{\bibfnamefont{G.~S.} \bibnamefont{Lewis}} \bibnamefont{and}
  \bibinfo{author}{\bibfnamefont{H.~L.} \bibnamefont{Swinney}},
  \bibinfo{journal}{Phys. Rev. E} \textbf{\bibinfo{volume}{59}},
  \bibinfo{pages}{5457} (\bibinfo{year}{1999}).

\bibitem[{\citenamefont{{Chevillard} et~al.}(2005)\citenamefont{{Chevillard},
  {Roux}, {L{\'e}v{\^e}que}, {Mordant}, {Pinton}, and
  {Arn{\'e}odo}}}]{chevillard05}
\bibinfo{author}{\bibfnamefont{L.}~\bibnamefont{{Chevillard}}},
  \bibinfo{author}{\bibfnamefont{S.~G.} \bibnamefont{{Roux}}},
  \bibinfo{author}{\bibfnamefont{E.}~\bibnamefont{{L{\'e}v{\^e}que}}},
  \bibinfo{author}{\bibfnamefont{N.}~\bibnamefont{{Mordant}}},
  \bibinfo{author}{\bibfnamefont{J.-F.} \bibnamefont{{Pinton}}},
  \bibnamefont{and}
  \bibinfo{author}{\bibfnamefont{A.}~\bibnamefont{{Arn{\'e}odo}}},
  \bibinfo{journal}{Phys. Rev. Lett.} \textbf{\bibinfo{volume}{95}},
  \bibinfo{pages}{064501} (\bibinfo{year}{2005}).

\bibitem[{\citenamefont{Fineberg et~al.}(1993)\citenamefont{Fineberg, Lathrop,
  and Swinney}}]{fineburg93}
\bibinfo{author}{\bibfnamefont{J.}~\bibnamefont{Fineberg}},
  \bibinfo{author}{\bibfnamefont{D.~P.} \bibnamefont{Lathrop}},
  \bibnamefont{and} \bibinfo{author}{\bibfnamefont{H.~L.}
  \bibnamefont{Swinney}}, in \emph{\bibinfo{booktitle}{Turbulence in Spatially
  Extended Systems}}, edited by
  \bibinfo{editor}{\bibfnamefont{R.}~\bibnamefont{Benzi}},
  \bibinfo{editor}{\bibfnamefont{C.}~\bibnamefont{Basdevant}},
  \bibnamefont{and} \bibinfo{editor}{\bibfnamefont{S.}~\bibnamefont{Ciliberto}}
  (\bibinfo{publisher}{Nova Science Publishers}, \bibinfo{year}{1993}).

\bibitem[{\citenamefont{{Richardson}}(1926)}]{richardson26}
\bibinfo{author}{\bibfnamefont{L.~F.} \bibnamefont{{Richardson}}},
  \bibinfo{journal}{Proc. R. Soc. London. A} \textbf{\bibinfo{volume}{110}},
  \bibinfo{pages}{709} (\bibinfo{year}{1926}).

\end{thebibliography}
